\documentclass[iop]{emulateapj}
\usepackage{times,natbib,graphicx,amsmath,multirow,xspace,ulem}
\usepackage{xcolor}
\usepackage{lipsum}
\usepackage{threeparttable}

\def \mnras {MNRAS}
\def \apj {ApJ}

\def \aap {A\&A}

\def \atel {ATel}

\begin{document}
\newcommand{\risco}{R_{\mathrm{ISCO}}}
\newcommand{\flux}{\mathrm{erg~cm}^{-2}~\mathrm{s}^{-1}}
\newcommand{\dist}{(D/5.8~\mathrm{kpc})^2}
\newcommand{\lum}{\mathrm{erg~s}^{-1}}
\newcommand{\vdag}{(v)^\dagger}
\newcommand\aastex{AAS\TeX}
\newcommand\latex{La\TeX}
\newcommand{\cgs}{erg cm$^{-2}$ s$^{-1}$}
\newcommand{\rosat}{{\it ROSAT}}
\newcommand{\sax}{{\it BeppoSAX}}
\newcommand{\asca}{{\it ASCA}}
\newcommand{\xmm}{{\it XMM-Newton}}
\newcommand{\xte}{{\it RXTE}}
\newcommand{\exosat}{{\it EXOSAT}}
\newcommand{\einstein}{{\it Einstein}}
\newcommand{\integral}{{\it INTEGRAL}}
\newcommand{\swift}{{\it SWIFT}}
\newcommand{\nustar}{{\it NuSTAR}}
\newcommand{\chandra}{{\it Chandra}}
\newcommand{\temna}{{\it Temna}}
\newcommand{\tnm}{\tablenotemark}
\newcommand{\nd}{\nodata}
\newcommand{\nh}{$N_{\rm H}\,$}
\newcommand{\mcl}{\multicolumn}
\newcommand{\cnts}{\mathrm{c~s}^{-1}}

\def\chiq{$\chi^2$}
\def\bb{{\sc bb}}
\def\nh{{$N_{\rm H}$}}
\def\I{{\it INTEGRAL}}
 \def\source{1RXS~J180408.9$-$342058}
\def\be{\begin{equation}}
\def\ee{\end{equation}}
\shorttitle{\aastex\ sample article}
\shortauthors{Fiocchi et al.}


\title{Quasi-simultaneous \integral\/, \swift\/, and \nustar\/  observations of the new X-ray {\it clocked} burster 1RXS~J180408.9-342058}


\author{M. Fiocchi$^1$}
\author{A. Bazzano$^1$}
\author{G. Bruni$^1$}
\author{R. Ludlam$^{2,\dagger}$}
\author{L. Natalucci$^1$}
\author{F. Onori$^1$}
\author{P. Ubertini$^1$}

\affiliation{$^1$Istituto di Astrofisica e Planetologia Spaziali -INAF-  Via Fosso del Cavaliere 100,  Roma, I-00133, Italy}
\affiliation{$^2$Cahill Center for Astronomy and Astrophysics, California Institute of Technology, Pasadena, CA 91125, USA\\ $^{\dagger}$ Einstein Fellow}

\begin{abstract}
We report the quasi-simultaneous \integral\/, \swift\/, and \nustar\/ observations showing spectral state transitions in the neutron star low-mass X-ray binary \source\/ during its 2015 outburst.  
We present results of the analysis of high-quality broad energy band (0.8--200~keV) data in three different spectral states: high/soft, low/very-hard, and transitional state.
The broad band spectra can be described in general as the sum of thermal Comptonization and reflection due to illumination of an optically-thick accretion disc. 
During the high/soft state, blackbody emission is generated from the accretion disc and the surface of the neutron star. This emission, measured at a temperature of $kT_{bb}\sim1.2$~keV, is then Comptonized by a thick corona with an electron temperature of $\sim2.5$~keV. 
For the transitional and low/very-hard state, the spectra are successfully explained with emission from a double Comptonizing corona. The first component is described by thermal Comptonization of seed disc/neutron-star photons ($kT_{bb}\sim1.2$~keV) by a cold corona cloud with $kT_e\sim8-10$ keV, while the second one originates from lower temperature blackbody photons ($kT_{bb}\leq 0.1$~keV) Comptonized by a hot corona ($kT_e\sim35$ keV).

Finally, from \nustar\/ observations, there is evidence that the source is a new {\it clocked} burster. 
The average time between two successive X-ray bursts corresponds to $\sim$ 7.9 ks and $\sim$ 4.0 ks when the persistent emission decreases by a factor $\sim$ 2, moving from very hard to transitional  state.\\
The accretion rate ($\sim4\times10^{-9}M_{\sun}/yr$) and the decay time of the X-ray bursts longer than $\sim$30 s suggest that the thermonuclear emission is due to mixed H/He burning triggered by thermally unstable He ignition.

\end{abstract}


\section{Introduction} \label{sec:intro}

Low mass X-ray binaries (LMXB) are double star systems consisting of a low-mass object in orbit around a black hole or neutron star.
When emission from an X-ray burst is observed the compact object can immediately be classified as a neutron star (NS).
According to our present understanding, the X-ray emission in X-ray Bursters comes from the release of gravitational potential energy  from accretion processes onto the neutron star. 
The X-ray spectra are generally described as the sum of a soft thermal component originating from the accretion disc and/or neutron star surface, a hard X-ray component arising from Inverse Compton scattering of soft thermal photons in a hot electron corona and a reflection component originated from the scattering of the Comptonized photons by the surface of the accretion disc. The emission of NS LMXBs shows two spectral states.
In the high/soft state the disc is generally thought to extend down to the neutron star and it is responsible for the thermal emission, being the disc and neutron star the source of the soft seed photons for Comptonization in the coronal region.
In the low/hard state, the disc is believed to recede from neutron star resulting in a weakened thermal component while the inner parts of a hot accretion flow produce hard X-ray emission.
The spectral transitions are accompanied by luminosity variations and are generally explained in the framework of this truncated accretion disc model (see Done et al.\ 2007).
In the low/hard state, the neutron star surface and disc emission are much weaker than in the high/soft state and the inner cavity of the accretion disc could be replaced by a hot geometrically thick accretion flow. 
Comptonization of the corona electrons interacting with the soft photons from disc/neutron star and/or from photons internally generated by synchrotron of the coronal electrons produce hard X-ray emission (Veledina, Poutaten, \& Vurm, 2013).
In both spectral states the disc receives photons from the corona and produces the reflection and reprocessing features. 
The truncated disc scenario explains qualitatively many of the observed properties of X-ray binaries. However, the geometry and nature of the corona in all spectral states are still uncertain. X-ray emission may arise from different regions or from different physical processes, such as a multi-zone structure with an inhomogeneous hot accretion flow (Veledina et al.\ 2013) or a jet contributing to high energy emission (Markoff, Nowak, \& Wilms, 2005), although at X-ray wavelengths the jet emission is unlikely to be dominant (Malzac, Belmont, \& Fabian, 2009). 
Recently, there have been many attempts to accurately derive information on the innermost region surrounding NSs in LMXBs, such as the physical properties of the emission region of the reflection components and the geometry of the system.  The use of broad band spectra together with  improved methods for self consistent spectral modelling to simultaneously fit the iron line profile and the Compton hump have proved to be a useful tool in this framework (Di Salvo et al.\ 2019, Chiang et al.\ 2016). In particular, the inferred values for the reflection amplitude give important information about the geometry of the emission region. In general, low values for the reflection amplitude have been found for NS LMXBs, lying between $0.2-0.3$ (Matranga et al.\ 2017a, 2017b; Di Salvo et al.\ 2015; Pintore et al.\ 2015), where a value of 0.3 is indicative of a spherical corona (Matranga et al.\ 2017b). In some cases, even smaller reflection fractions are observed. For example, Mazzola et al.\ (2018) reported a reflection amplitude of 0.05 in the {\it XMM-Newton} and {\it INTEGRAL} spectra of 4U 1702$-$49, which could indicate a different geometry with respect a spherical corona (but see also the analysis of Iaria et al.\ 2016).

\source\/ was first detected in 1990 by the \rosat\/ satellite (Voges et al.\ 1999) and underwent a faint outburst  in 2012 that was caught by \integral\/ and \swift\/ (Chenevez et al.\ 2012).
During \integral\/ Galactic Bulge monitoring observations, JEM-X instruments detected a Type-I thermonuclear burst which classified the compact object in this system as a neutron star.  Assuming that this burst was Eddington limited, Chenevez et al.\ (2012) determined an upper limit on its distance to be $d\leq5.8$ kpc.
Follow-up \swift\/ observations revealed that \source\/ returned to quiescence (Kaur \& Heinke, 2012) and it remained until 2015 January 20, when both \swift\//BAT and {\it MAXI}/GSC detected a new outburst (Krimm et al.\ 2015a,b; Negoro et al.\ 2015). 
\integral\/ observations of the Galactic Center on 2015 February 16-17 detected this source in a hard state, with emission up to 100~keV (Boissay et al. 2015).  
The X-ray spectrum was described by a simple absorbed power-law model with a photon index of $\Gamma\sim$1.1 (Degenaar et al. 2015).

On April 3, the hard X-ray flux dropped and the soft X-ray emission increased, indicating a transition from a low/hard to high/soft state (Degenaar et al.\ 2015).
Baglio et al.\ (2016) reported on a detailed NIR/optical/UV study of \source\/ during both the low/hard state (MJD 57079) and high/soft state (MJD 57136). 
The optical spectrum showed He I emission lines with a lack of H emission lines that are typically observed in LMXBs, suggesting an ultra-compact nature for this transient source. 
By combining the mass accretion rate with theoretical evolutionary tracks for a He white dwarf, Baglio et al.\ (2016) reported on a tentative orbital period of $\sim$ 40 min. 
Based on the multi-wavelength information, Degenaar et al.\ (2016) concluded that orbital periods of $\sim$ 1-3 hr are not ruled out.
Furthermore, Baglio et al.\ (2016) studied the spectral energy distribution evolution, showing the emission was consistent with a simple thermal component during the soft state, while the presence of a tail in the NIR in addition to the thermal component during the hard state likely indicated a transient jet. 
 
Detailed studies of the very hard and soft states of \source\/ were performed with \nustar\/ observations performed during the 2015 outburst (Ludlam et al.\ 2016; Degenaar et al.\ 2016).
Ludlam et al.\ (2016) reported on the source in the very hard spectral state, showing multiple reflection features (Fe K$_{\alpha}$ detected with \nustar\/; N VII, O VII, and O VIII detected with \xmm\//RGS) from different ionization zones.
Using relativistic reflection models they measured an inner disc radius $\leq22.2$ km and an inclination of the system between $18^{\circ}-29^{\circ}$.
Degenaar et al.\ (2016) presented \nustar\/, \swift\/, and \chandra\/ observations obtained around the peak brightness of this outburst ($L_{\mathrm{0.5-10~keV}}$$\simeq$$(2-3)\times10^{37}~\dist~\lum$) while the source was in the soft spectral state. 
The \nustar\ data still showed a relativistically broadened Fe K emission line that indicated that the inner edge of the accretion disc extending down to $\sim11-17$\ km from the neutron star. 
Additionally, the inclination inferred from reflection modeling ($i\simeq27^{\circ}-35^{\circ}$) agreed with the results obtained while the source was in the hard spectral state.   

From all \swift\/XRT observations obtained during the 2015 outburst of \source\/, Parikh et al.\ (2017b) found that during the low/hard state this source showed a photon index of $\Gamma\sim1$. This is significantly lower than typical values of $\Gamma=1.5-2.0$ for such systems in the hard state. The mechanism for producing such a hard spectrum remained unclear.
Wijnands et al.\ (2017) used two XMM-Newton observations to study the rapid variability properties of \source\/: one obtained early during the outburst in March 6 (see also Ludlam et al.\ 2016) and one just before the transition to the soft state on April 1.
These authors reported the source exhibited an unusually strong noise component, particularly during the first {\it XMM-Newton} observation. At X-ray luminosities similar to those observed in \source\/, they showed its timing properties are different with respect to what are commonly observed in the canonical hard state of a neutron star.

Parikh et al.\ (2017a) reported on the \swift\/ and \xmm\/ observations while the source was in the quiescence state. 
The X-ray spectra were dominated by a thermal component arising from the stellar surface temperature that decayed after the end of outburst.
Good spectral fits were obtained from these data by adding a power-law component in addition to the existing thermal emission. The power-law contributed 30\% of the total unabsorbed flux in 0.5--10~keV energy range. The origin of this component is unknown.

 Recently, Gusinskaia et al.\ (2017) presented an EVLA monitoring study of the source
through the 2015 outburst and back to quiescence for a total of 6 epochs. 
The enhanced sensitivity of the EVLA allowed them to follow the evolution of the radio counterpart at X-band (8-12 GHz) from the very hard X-ray state ($232 \pm 4\ \mu Jy$) down to the soft X-ray state ($19 \pm 4\ \mu Jy$). 
When source faded to quiescence, a flux density upper limit of  $\leq 13\ \mu Jy$ was measured. 
Moreover, the wide bandwidth of the EVLA allowed them to perform a spectral analysis, 
resulting in a positive spectral index ($0.12\pm0.18$) during the very hard state. 
The positive spectral index implied that the emission came from an optically thick jet, which agreed with the expectations for sources in hard X-ray states. 
However, the larger uncertainties of the soft-state measurements did not allow a proper 
estimation of the spectral index. Thus, they were not able to discern a change from optically thick to optically thin emission. 
These radio observations were performed with the EVLA in B and BnA configurations and resulted in an angular resolution of 1.25$^{\prime\prime}$ and 0.6$^{\prime\prime}$, respectively. As a result, 
the source was resolved from the nearby extended radio galaxy NVSS J180414−342238 (Condon et al.\ 1998). 
Previous NVSS observations of this region, with an angular resolution of 45$^{\prime\prime}$, would not have allowed a clear detection despite the possible presence of a very hard state. 
Gusinskaia et al.\ (2017) conclude that the weak radio emission seen in the soft state could still be residual emission from the very hard state, thus jet quenching could be even deeper than what was measured (more than one order of magnitude).

\begin{deluxetable*}{cccc}[h]
\tablecaption{Observation log \label{tab:jou}}
\tablecolumns{4}
\tablewidth{0pt}
\tablehead{
\colhead{\#}&
\colhead{Instrument} &
\colhead{Start Date}&
\colhead{Duration} \\
\colhead{} & 
\colhead{} & 
\colhead{(UTC)} & 
\colhead{(ks)}
}
\startdata
1&{\it INTEGRAL}/IBIS   & 2015-02-16 12:34:50	 &  5929.9\\
2&{\it Swift}/XRT       & 2015-02-17 17:21:58 & 0.960 \\
3&		  & 2015-02-24 05:58:59 & 1.982 \\
4&       & 2015-02-26 02:27:59 & 1.848 \\
5&       & 2015-02-28 09:00:59 & 0.555 \\
6&       & 2015-03-02 21:32:06 & 1.099 \\
7&       & 2015-03-04 15:05:59 & 2.169 \\
8&      & 2015-03-09 02:04:59 & 1.734 \\
9&       & 2015-03-10 13:26:59 & 1.489 \\
10&       & 2015-03-13 00:11:59 & 0.595 \\
11&       & 2015-03-16 06:45:59 & 0.980 \\
12&       & 2015-03-23 06:12:52 & 0.200 \\
13&       & 2015-03-24 22:33:58 & 0.590 \\
14&      & 2015-04-01 16:52:59 & 1.858 \\
15&       & 2015-04-14 13:00:13 & 1.000 \\
16&       & 2015-04-15 09:46:59 & 1.839 \\
17&{\it NuSTAR}/(FPMA,FPMB)  & 2015-03-05 09:21:07 & 80.096 \\
18&  & 2015-04-01 16:16:07 & 56.996 \\  
19&  & 2015-04-14 12:11:07 & 45.596\\
\enddata
\end{deluxetable*}

\section{Observations and Data Analysis}
\label{observations}
To understand the spectral and timing properties of \source\/ we analyze the quasi-simultaneous data in a homogeneous manner over a very broad energy band ($0.8-200$ keV). We report in 
Table \ref{tab:jou} the observation log of the source performed with {\it INTEGRAL}/IBIS (Winkler et al.\ 2003), {\it SWIFT}/XRT (Gehrels et al.\ 2004), and {\it NuSTAR}/FPMA-FPMB (Harrison et al.\ 2013) instruments. 

The {\it INTEGRAL}/IBIS (Ubertini et al.\ 2003) data are processed using the standard Off-line Scientific Analysis (OSA v10.2) software released by the \integral\/ Scientific Data Centre (Courvoisier et al.\ 2003).

The \swift\//XRT (Barrow et al.\ 2005) data are processed using standard tools incorporated in \textsc{heasoft} (v6.19). 
Observations were taken in windowed timing mode. We used a box region with a length of $70''$ to extract source events and another region of the CCD away the source for background events. 
When the XRT count rate was $>$100$~\cnts$ we used a box-box region with a length of $70''$  and an exclusion of the inner of $2''$ to avoid pile-up effect (following the procedure in Romano et al.\ 2006). 
\\
The \nustar\//FPMA and FPMB instruments data are processed following standard analysis, performed with \textsc{nustardas} included in \textsc{heasoft} (v.6.19).
We used \textsc{nuproducts} to create light curves and spectra for the FPMA and FPMB, considering  a circular region with a radius of $60''$  to extract source events and a region of the same size away the source to extract background events. 
\\

For spectral fitting of each spectral state, we allow a constant to be free with respect to the FPMA data, for IBIS, XRT, and FPMB instruments. 
We use XSPEC v.\ 12.9.0n and data available in the following energy range: $0.8-9.0$ keV for \swift\//XRT, $3.5-50.0$ keV for \nustar\//FPMA and FPMB, and 
$22.0-200.0$ keV for {\it INTEGRAL}/IBIS instrument. 
The \swift\//XRT data below 0.8~keV are not included because \xmm\//RGS observations showed multiple emission lines (N VII, O VII and O VIII), which cannot be resolved using the XRT data.

\begin{table*}[h]
\caption{Results from fitting \integral\/, \nustar\/, and \swift\/ spectral data of the persistent emission.}
\begin{tabular*}{0.99\textwidth}{@{\extracolsep{\fill}}l l l c c c  }
\hline
\multicolumn{3}{c}{Parameter (unit)} & Very hard State&Transitional State &Soft State \\
\hline
   && &&&  \\
\multicolumn{3}{c}{Exposure time XRT (s) }\dotfill & 13170& 1837 &2813\\
\multicolumn{3}{c}{Exposure time  FPMA (ks)}\dotfill & 35.43& 24.77&20.23\\
\multicolumn{3}{c}{Exposure time  FPMB (ks)} \dotfill & 35.56&25.00 &20.51\\
\multicolumn{3}{c}{Exposure time  IBIS (ks)} \dotfill &438.9 &16.42 &278.6\\
   &&  && & \\
\hline
& &  &&   & \\
Component 	& \#					    & Parameter           & Model 2                 & Model 2          & Model 1\\
 &  & &  &&  \\
            \hline
 &  & &  &&  \\
CONSTANT    &  1   &$C$  (FPMB) \dotfill 				        & $1.004 \pm 0.002$     & $1.010\pm 0.008$  		& $0.996\pm 0.004$ \\
            &  2   &$C$  (XRT) \dotfill 				        & $1.03 \pm 0.05$       & $0.91 \pm 0.08$   	 	& $1.08 \pm 0.04$  \\
            &  3   &$C$  (IBIS) \dotfill 				        & $1.1 \pm 0.5$         & $1.2 \pm 0.5$      		& $1.1 \pm 0.2$  \\
 &  & &  &&  \\
            \hline
 &  & &  &&  \\
TBABS       &   4  &$N_H$ ($\times10^{22}~\mathrm{cm}^{-2}$) \dotfill 	& $0.3 \pm 0.1$ 	                            & $0.3 \pm 0.2$             &	$0.3 \pm 0.1$  \\
  &  & &  &&  \\
            \hline
 &  & &  &&  \\
DISKBB      &   5  &$kT_{\mathrm{in}}$  (keV) \dotfill 		    &           ...                      &       ...      &$1.15\pm0.02$\\
            &   6  &$N_{\mathrm{DISKBB}}$  $(\mathrm{R_{in}^{km}}/D_{10~\mathrm{kpc}})^2 \cos i$ \dotfill       & ...  & ...  &$21\pm6$ \\
  &  & &  &&  \\
            \hline
 &  & &  &&  \\
NTHCOMP   &  7   &$kT_{\mathrm{e}}$  (keV) \dotfill		            &      $10\pm 2$             &  	 $8\pm2$                        &$2.50\pm 0.05$       \\
            &  8  &$kT_{\mathrm{bb}}$  (keV)  \dotfill		        &    $1.7_{-0.3}^{+0.5} $   &    $1.7\pm0.8 $                   &$1.2\pm0.2$    \\
        &  9   &$\Gamma$  \dotfill  				                &   1.9$\pm0.2$             &     $1.9\pm0.3 $ 	                &$2.2\pm 0.2$       \\
        &  10   &$N_{\mathrm{nthcomp}}$($\times10^{-2}$)\dotfill	&   $8\pm 2$   		        &$12\pm 4$      		            &$48\pm10$     \\
RDBLUR  &11 &$Betor10$  \dotfill  				                    & $-1.8\pm0.8$              &$-1.6\pm1.5$                       &$-6\pm 5$\\
       &  12& R$_{in}$ (R$_{G}$)\dotfill 	                        &  $13_{-8}^{+45}$          &  $7_{-2}^{+17}$                   &$6\pm1$\\
        & 13&$i$  ($^{\circ}$) \dotfill 			                &$51_{-18}^{+14}$           & $53_{-28}^{+10}$                  &$40\pm16 $ \\
XILCONV &14 &$refl\_frac$ \dotfill 				                    & $\leq$0.2  	            &   $0.3\pm $0.1       	       	    &$0.9_{-0.5}^{+1.0} $\\
       & 15 & $A_{\mathrm{Fe}}$  ($\times$Solar) \dotfill 	        & $0.5\pm0.4$               & $0.5\pm0.4$ 	                    &$0.8\pm0.4$\\
        &   16  &$log \xi$  ($\mathrm{erg~cm~s}^{-1}$) \dotfill     &  $1.3\pm0.4$              &    $1.9\pm0.6$                    &$1.8_{-1.5}^{+0.2}$ \\
 &  & &  &&  \\
            \hline
 &  & &  &&  \\
NTHCOMP   &  17  &$kT_{\mathrm{e}}$  (keV) \dotfill		            &    35$\pm8$	        &   $34\pm8$ 	            &...\\      
           & 18  &$kT_{\mathrm{bb}}$  (keV)  \dotfill		        &  $\leq $0.1           &   $\leq $0.1              &...    \\  
           &  19 &$\Gamma$  \dotfill  				                &   $1.9\pm0.2 $  	    &    $1.7\pm0.2 $	        &...    \\  
           &  20 &$N_{\mathrm{nthcomp}}$($\times10^{-2}$)\dotfill	& 3$\pm2$		        &	$10\pm3$	            &...  \\    
RDBLUR & 21 &$Betor10$  \dotfill  				                    & = \#11  	            & = \#11                    &...\\      
       & 22 & R$_{in}$ (R$_{G}$)\dotfill 	                        & = \#12                & = \#12	                &...\\      
       & 23 &$i$  ($^{\circ}$) \dotfill 			                & = \#13                & = \#13                    &  ... \\   
XILCONV &24  &$refl\_frac$ \dotfill 				                &   1.7$\pm0.5$         &  $0.4\pm0.3$              &...\\      
       & 25 & $A_{\mathrm{Fe}}$  ($\times$Solar) \dotfill 	        & = \#15 	            & = \#15 	                &...\\      
        & 26&$log \xi$  ($\mathrm{erg~cm~s}^{-1}$) \dotfill         &  	$1.8\pm0.5$         &  $3.1\pm0.8$    	        & ... \\    
 &  & &  &&  \\
            \hline
 &  & &  &&  \\
& &$F_{\mathrm{0.8-200~keV}} (\times10^{-10}~\mathrm{erg}~\mathrm{cm}^{-2}\ \mathrm{s}^{-1})$\dotfill  &		$45\pm13$	& $55\pm13$  & $56\pm5$	\\
& &$L_{\mathrm{0.8-200~keV}} (\times10^{37}~\mathrm{erg}~\mathrm{s}^{-1})$\dotfill  &		$1.8\pm0.6$	& $2.2\pm0.6$  & $2.6\pm$0.2	\\

 &    &&  &&\\
& &$\chi^2_{red}$ (dof) \dotfill  			&  0.89 (3045) & 0.90 (2633)	&0.91(1602)\\
  &  && & &  \\
\hline
\end{tabular*}
\label{tab:reflspec}
\begin{tablenotes}
\item

We fixed  $R_{\mathrm{out}}$$=$$500~\risco$. All reflection fraction values are set to less than 0 to consider only the reflected component, we report the absolute value of this component for clarity. 
Fluxes are computed in $0.8-200$ keV energy range and they are unabsorbed fluxes. The luminosity is given assuming a distance of 5.8 kpc to the source.
The constant multiplication factor was fixed to $C=1$ for the \nustar/FPMA data and 
left free for the other instruments. 
Quoted errors reflect 1$\sigma$ confidence levels.

\end{tablenotes}
\end{table*}

\section{Results}
\label{analysis}
This work  reports on all three spectral states in a broad energy range ($0.8-200.0$ keV) and,  in particular, the only data set that has high quality broad band X-ray data available in the peculiar spectral/timing state, i.e., the so called very hard state by Parikh et al.\ (2017) and Wijnands (2017). 
Both the persistent and X-ray burst emission have been analyzed in this work. To study the persistent emission, 
we extracted \integral/IBIS, \swift/XRT, \nustar/FPMA and \nustar/FPMB   
light curves and spectra excluding the times of Type-I X-ray bursts as reported
 in Section~\ref{subsec:steady}. Timing and spectral analysis of the Type-I X-ray bursts is in Section~\ref{subsec:bursts}.\\

\subsection{The persistent emission}
\label{subsec:steady}

\begin{figure*}[t]
\includegraphics[angle=-90,width=18.0cm]{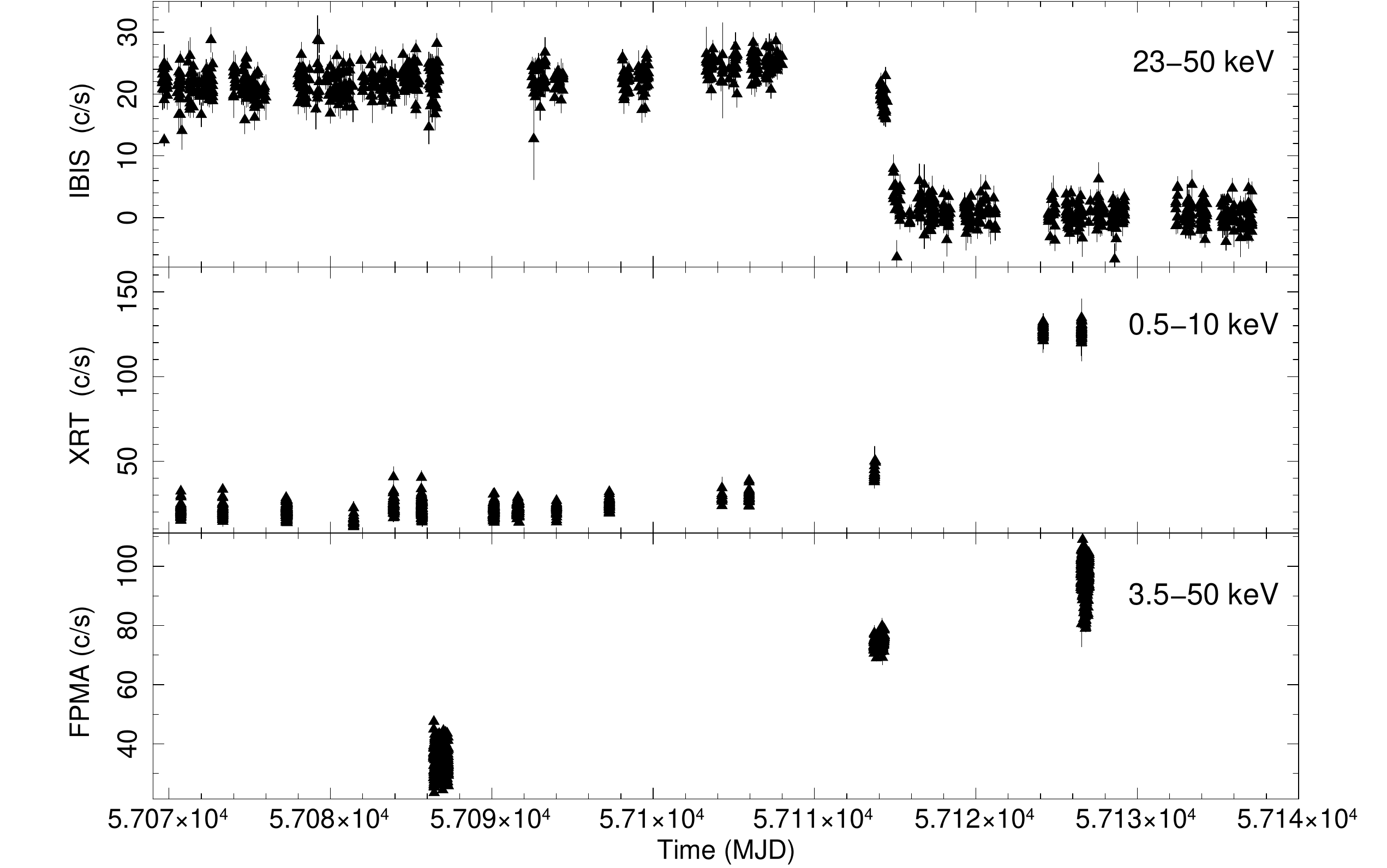}\\
\caption{Top panel shows the IBIS light curve in the $23-50$~keV energy band from 2015 February 16 (57069 MJD) to 2015 April (57137 MJD). Middle panel reports on the XRT light curve in the $0.5-10$~keV 
energy band. 
The bottom panel shows the FPMA light curve in the $3.5-50$~keV energy band.\label{fig:lc}}
\end{figure*}

Fig.\ \ref{fig:lc} (top panel) shows the IBIS light curve in the \mbox{$23-50$~keV} energy band from 2015 February 16 (57069~MJD) to 2015 April (57137~MJD). 
This chosen energy range ($0.8-200$~keV) is the one that best represents the X-ray emission simultaneously both in the soft and hard state. 
From the beginning of the observation up to $\sim$57110 MJD the source was in 
a very hard X-ray state ($F_{\mathrm{23-50~keV}}\sim10^{-9}$ \cgs).
Starting from $\sim\/$57114 MJD,  \source\/ entered into a 
transitional state as it passed from the very hard to the soft state. It stayed in the soft state up to the end of the \integral\/ observation ($F_{\mathrm{23-50~keV}}\sim 2\times 10^{-11}$ \cgs) during the time interval $57116-57137$ MJD.
The middle panel of Fig.\ \ref{fig:lc} reports on the XRT light curve in the $0.5-10$~keV energy band showing a gradual increase of the soft X-ray emission, confirming the transition of the source from the hard to the soft spectral state. 
The bottom panel of Fig. \ref{fig:lc} shows the FPMA light curve in the $3.5-50$~keV energy band. 
\nustar\/ has observed this source in three occasions: the first during the very hard state, the second during the transitional state, and the third when the source moved to the soft state.

Simultaneous data from three high energy instruments allow for the spectral evolution to be followed during the state transitions over a very broad energy band. 
To this aim we study the spectral behavior in a uniform manner separately for three different states:
\begin{itemize}
\item
 Very hard state: the \integral\/, \swift\/, and \nustar\/ observations performed between MJD 57069 and MJD 57110.
\item
 Transitional state: the \integral\/, \swift\/, and \nustar\/ observations performed on MJD 57114.
\item
 Soft state: the \integral\/, \swift\/, and \nustar\/ observations performed between MJD 57116 and MJD 57137.
\end{itemize}

Previous analyses of the \nustar\/ data (Ludlam 2016; Degenaar 2016) showed a complex spectrum from the steady emission during two spectral states (very hard and soft)
consisting of Comptonization and reflection components with disc emission seen in the soft state only.
High quality data from transitional state are published in this work for the first time, but see also Marino et al., 2019.\\

\begin{figure}[h]
\includegraphics[angle=-90,scale=0.36]{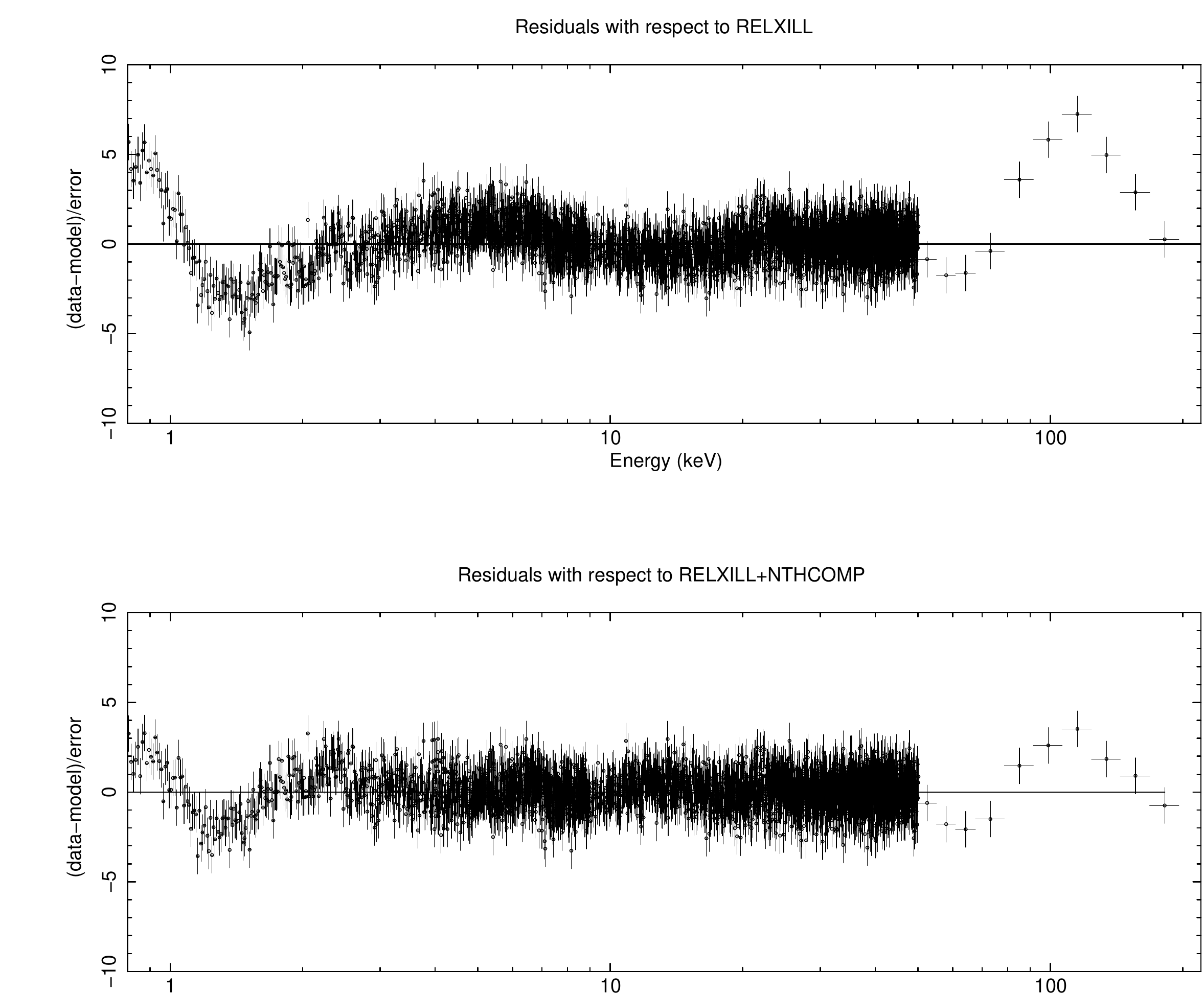}\\
\caption{Residuals in units of sigma for the low/hard state spectra with respect to the model  {\scriptsize{RELXILL}} (top panel) and {\scriptsize{RELXILL+NTHCOMP}}
 (bottom panel)  \label{fig:res}}
\end{figure}

\subsubsection{The Very Hard State}

To build the average broad energy band spectrum ($0.8-200.0$~keV) for a very hard state, we use the \nustar\/ data mentioned above, the average spectrum with the combined \swift\/ observations 
from 2 to 13 (see Table 1), and the \integral\/ data. To fit these data we use the self-consistent model for X-ray reflection for a cutoff power-law irradiating an accretion disc, modeled in XSPEC by {\scriptsize{RELXILL}} (Garcia et al.\ 2014). The {\scriptsize{RELXILL}} model parameters are as follows:
the emissivity $q_1$ between $R_{in}$ and $R_{break}$,  the emissivity $q_2$ between $R_{break}$ and $R_{out}$, break radius between the two emissivities $R_{break}$,
inner and outer radius of the accretion disc $R_{in}$, $R_{out}$, spin parameter $a$,
inclination of the disc $i$, ionization of the accretion disc $\xi$, the iron abundance of the material in the accretion disc $A_{Fe}$, the high energy cutoff of the incident power law  $E_{cut-refl}$, the reflection factor $refl\_frac$, power law index of the incident spectrum $\Gamma$ and normalization $N_{refl}$.

We begin with fitting the spectrum using this model proposed by Ludlam et al.\ (2016), which works well in the range $3.5-50$~keV, with all physical parameters being free in the fit, except for the fixed parameters $a=0$, $R_{\mathrm{in}}=1~\risco$ and  $R_{\mathrm{out}}=500~\risco$. 
When including the \integral\/ data covering energies up to 200~keV, this model gives a poor fit ($\chi^2_{\rm red} (d.o.f.)= 1.36 (3050)$). Figure \ref{fig:res} (top model) shows the residuals in the data with respect to the {\scriptsize{RELXILL}} model. Two broad features below 2 keV and near 100 keV can be seen. We try to add a thermal Comptonization component to the model and we obtain $\chi^2_{\rm red} (d.o.f.)= 1.10 (3046)$ with   significant residuals
at both soft and high energies with respect to this model, even though the addition of the Comptonized component provides a better overall fit (Figure \ref{fig:res}, bottom panel). 
Using this model, the addition of a soft disc black body component does  not improve the fit quality ($\chi^2_{\rm red} (d.o.f.)= 1.10 (3044)$).  Furthermore the best fit parameters result in an unphysical black body temperature of $\sim 5$~keV.

Considering this and the behaviour in the radio/IR band during the low/hard state (Baglio et al.\ 2016, Gusinskaia et al.\ 2017), we fit these data by adding a second emission component and its reflection.
To obtain the physical parameters of the primary continuum we use the {\scriptsize{XILCONV}} model. This model take into account Doppler and relativistic effects due to the fast motion of the matter in the disc, converting the {\scriptsize{XILLVER}} reflection table (which assumes a power law as illuminating spectrum) into a convolution for use with any continuum (Kolehmainem et al.\ 2011). It is parameterized by the reflection normalization $refl\_frac$, the  iron abundance of the material in the accretion disc $A_{Fe}$, the inclination of the disc $i$, ionization of the accretion disc $\xi$ and exponential cut-off energy $E_c$. The reflection fraction is set to negative values so that the model only accounts for the reflection component. We report the absolute value of this parameter in Table \ref{tab:reflspec}.  To take into account the relativistic smearing effects in the inner region of the accretion disc, we use  {\scriptsize{RDBLUR}} model, which is similar to {\scriptsize{DISKLINE}} (Fabian et al.\, 1989). This  convolution  model is parameterized by the index $Betor10$ of the power law, describing the emissivity, which scales as $r^{Betor10}$, the inner and outer radius $R_{in}$  and $R_{out}$ of the reflection region in units of gravitational radii ($GM/c^2$) and the inclination angle of the source $i$.
To include a Comptonization component as the incident emission, we fit data with the following model:
\begin{itemize}
\item {\scriptsize{CONST*TBABS*(NTHCOMP+RDBLUR*XILCONV*NTHCOMP)}} \\
 (Model 1)
\end{itemize} 

\noindent in which $inp\_type$ parameter was set to 1, indicating that the seed photons were emitted by the accretion disc. The outer radius was fixed at 500 gravitational radii ($R_g=GM/c^2$) as the fit was insensitive to its changes, the cutoff energy was fixed to be 2.7 times the temperature of the electrons (Egron et al.\ 2013).
All parameters in the {\scriptsize{NTHCOMP}} components were tied so that the resulting reflection emission is self-consistent with the illuminating continuum.
Using Model 1, we obtain a poor fit ($\chi^2_{\rm red} (d.o.f.)= 1.30 (3051)$) with residuals showing distortion at low energy and strong residuals at high energies.
We added a second thermal emission component {\scriptsize{NTHCOMP}}
with its reflection as following:  
\begin{itemize}
\item
{\scriptsize{CONST*TBABS*(NTHCOMP+RDBLUR*XILCONV*NTHCOMP+\\NTHCOMP+RDBLUR*XILCONV*NTHCOMP)}} \\ (Model 2)
\end{itemize}
We link the index $Betor10$, inclination angle $i$, the inner and outer radius $R_{in}$  and $R_{out}$ and the the  iron abundance $A_{Fe}$ of the first component to the same parameters of the second component. Conversely, we allow the ionization $\xi$, reflection $rel\_frac$ and normalization to change. 
This model improves the fit quality ($\chi^2_{\rm red} (d.o.f.)= 0.89 (3045)$). 
We add a soft disc black body component and we fit data with a model as 

\begin{itemize}
\item
{\scriptsize{CONST*TBABS*(NTHCOMP+RDBLUR*XILCONV*NTHCOMP\\+NTHCOMP+RDBLUR*XILCONV*NTHCOMP+DISKBB)}} \\ (Model 3)
\end{itemize}

Model 3 did not result in a substantial improvement in the fit of the data ($\chi^2_{\rm red} (d.o.f.)= 0.89 (3043)$). 
 The best fit parameters obtained with Model~2 are listed in Table \ref{tab:reflspec}  for the very hard state. The spectrum and model residuals in units of sigma are shown in Figure \ref{fig:spe} (top panel).

\subsubsection{The Transitional State}

During the transitional state, the hard X-ray flux $F_{\mathrm{23-50~keV}}\sim10^{-9}$ \cgs\ is similar to the flux during  the very hard state. 
For this reason, we try to fit these data with the same procedure used for the very hard state spectra. 
In the broad energy range $0.8-100$~keV the model proposed by Ludlam et al.\ (2016) with all physical parameters being free in the fit (with the exception of the aforementioned spin, inner disc radius, and outer disc radius) gives a poor fit ($\chi^2_{\rm red} (d.o.f.)= 1.28 (2627)$).  
Adding to the model a thermal Comptonization component, the fit quality reduces to $\chi^2_{\rm red} (d.o.f.) = 1.10 (2623)$. 

When we use Model 1 similar to the hard state, we obtain $\chi^2_{\rm red}$ (d.o.f.) of 1.5 (2639) while 
Model 2 well fits the transitional state data ($\chi^2_{\rm red} (d.o.f.) = 0.90 (2633)$). The {\scriptsize{DISKBB}} component is not required by the data  ($\chi^2_{\rm red} (d.o.f.) = 0.90 (2631)$).
The best fit parameters are listed in Table \ref{tab:reflspec} (Model 2) and the spectrum with residuals in sigma can be seen in the middle panel of Figure \ref{fig:spe}. 
  
\subsubsection{The Soft State}
\begin{figure}
\begin{center}
\includegraphics[angle=-90,width=8.7cm]{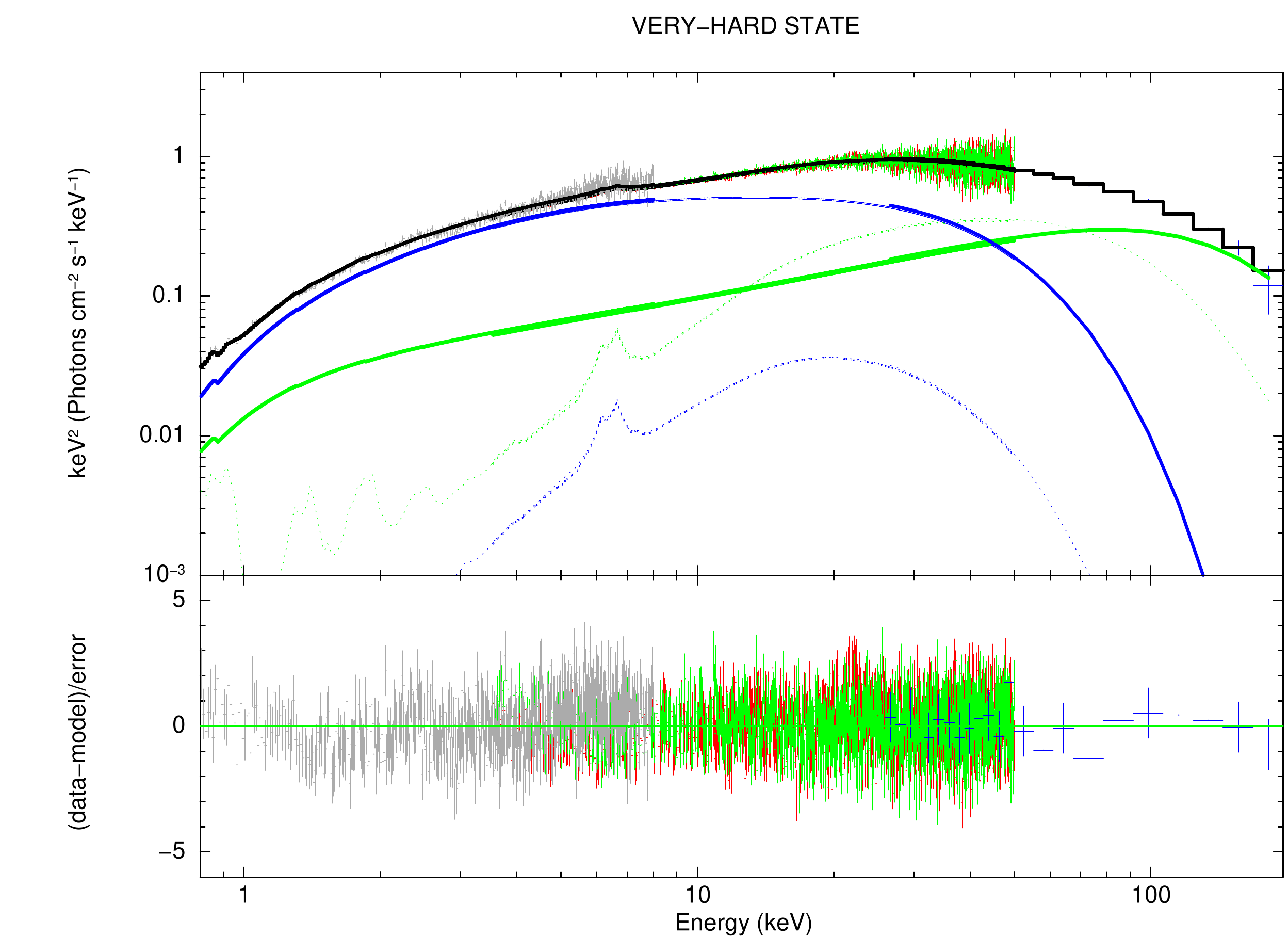}\\
\vspace{0.3cm}
\includegraphics[angle=-90,width=8.7cm]{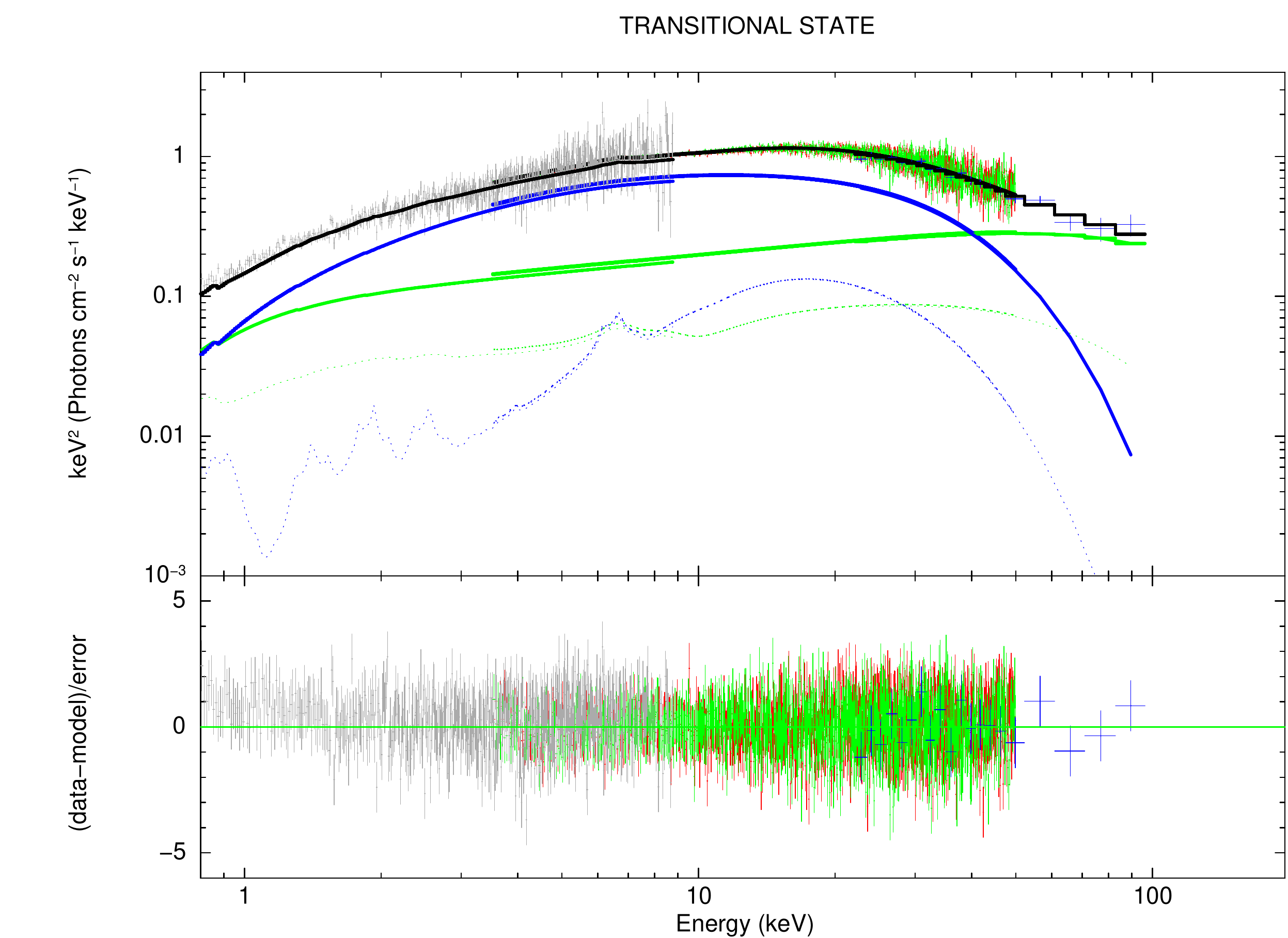}\\
\vspace{0.3cm}
\includegraphics[angle=-90,width=8.7cm]{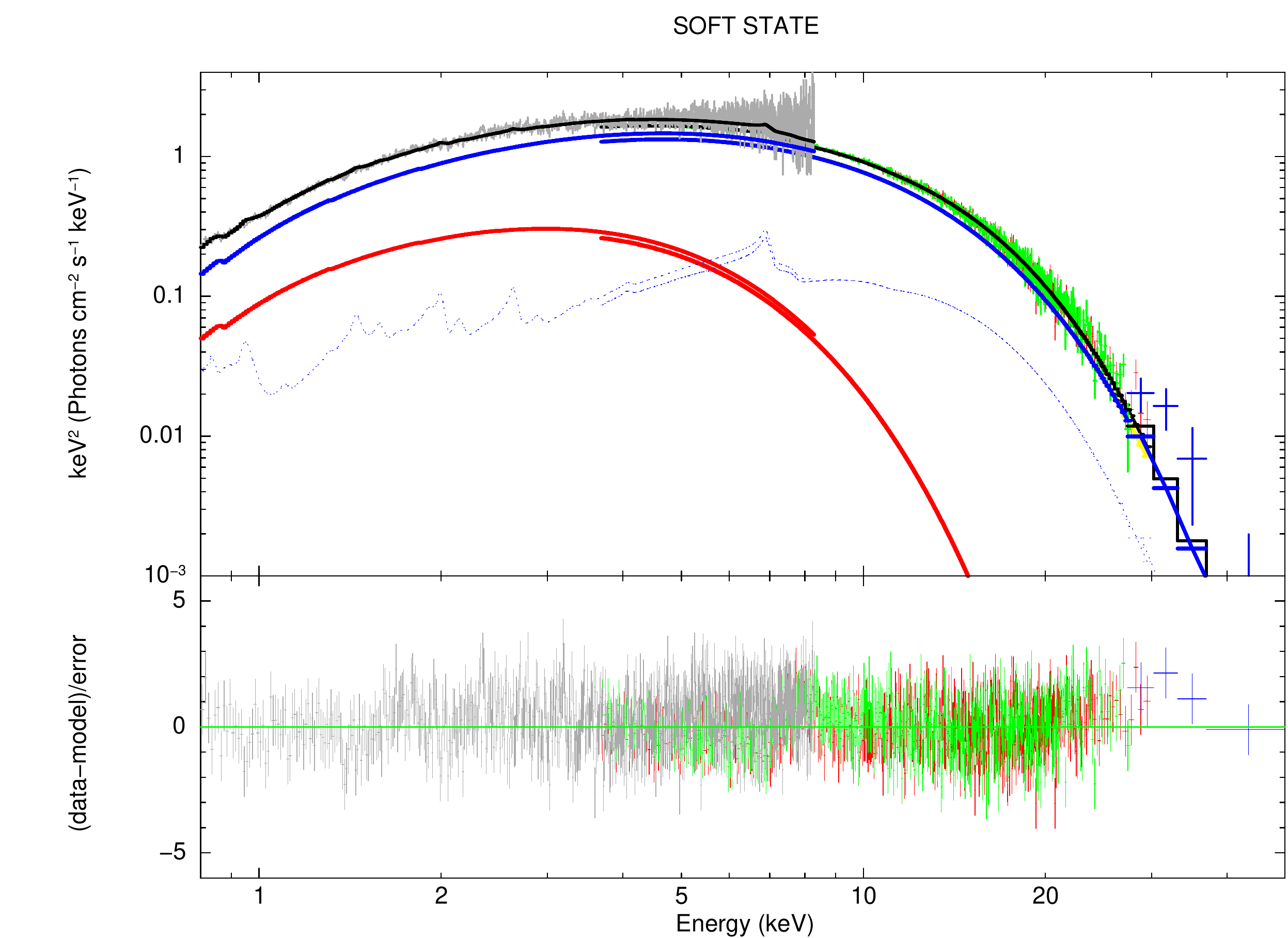}\\
\caption{\swift\//XRT, \nustar\//FPMA and FPMB, and {\it INTEGRAL}/IBIS unfolded spectra and residuals in sigma rebinned for visual clarity for the very hard state (top panel), transitional state (middle panel) and high/soft state (bottom panel). Light grey points correspond to XRT data, red and green points correspond to FPMA and FPMB data, and blue points are IBIS data. For the very hard and transitional state we plot the spectral components of Model 2 and Model 1 for soft state. Red lines correspond to DISKBB components, blue lines to NTHCOMP components and its reflection component (blue dotted), green lines correspond to second NTHCOMP components and its reflection (green dotted).
\label{fig:spe} }
\end{center}
\end{figure}

The \nustar\/ observation in the soft state and the \swift\/ observation number 15 (see Table 1) has previously been reported by Degenaar et al.\ (2016). The simplest model to describe these data consists of a thermal Comptonization component, a soft disc black body component,  and a reflection component.
To build an average broad band spectrum, we use the average \swift\//XRT spectrum of the observations number 15 and 16 (see Table 1) and the \integral\/ data extending the energy range up to $\sim50$~keV.  
Similarly as proposed by Degenaar et al.\ (2016), we fit the broad band  spectrum using a model including three emission component (a thermal Comptonization, a soft disc black body  and a model with reflection 
({\scriptsize{CONSTANT*TBABS*(DISKBB+NTHCOMP+RELXILL)}} in  \textsc{xspec}). 
All physical parameters are free in the fit with the exception of the following: $a=0$, $R_{\mathrm{in}}=1~\risco$, and $R_{\mathrm{out}}=400~\risco$ (values from Degenaar et al.\ 2016).
This model well describes broad band $0.8-50.0$~keV spectrum, with a  $\chi^2_{\rm red} (d.o.f.)= 0.83 (1600)$. 
Taking into account that the single Comptonization component described the high/soft state data well and, in order to study the behavior of spectral parameters, we then fit these data also with Model 1, similarly to what has been done for  the hard and transitional states. The best fit parameters are listed in Table \ref{tab:reflspec} and the spectrum for the soft state with residuals with respect to the model are shown in Figure \ref{fig:spe} (bottom panel).

\subsection{The Type I X-ray bursts }
\label{subsec:bursts}
\begin{figure*}
\begin{center}
\includegraphics[width=6.0cm,height=18cm,angle=-90]{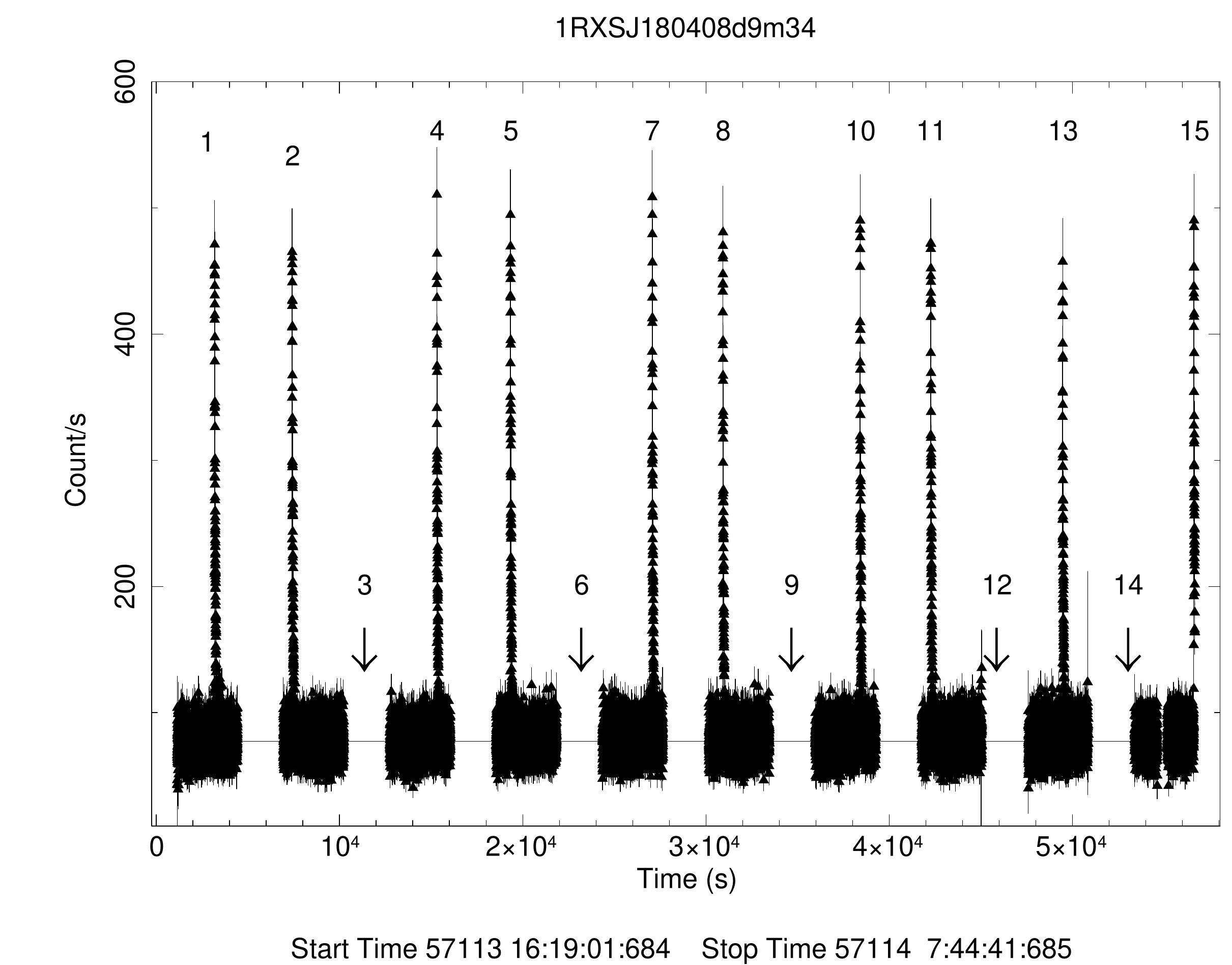}\\
\vspace{1cm}
\includegraphics[width=6.0cm,height=18cm,angle=-90]{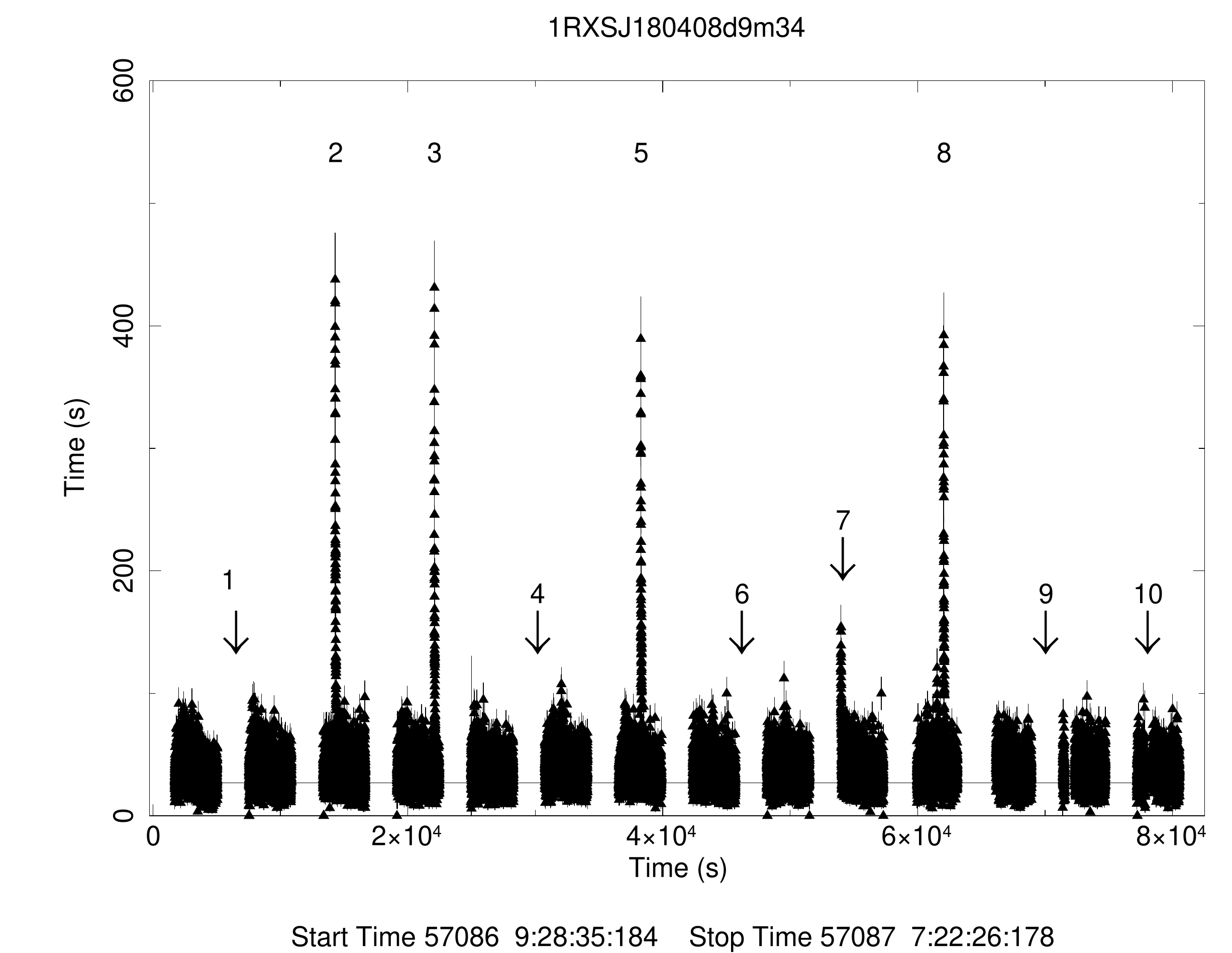}\\
\caption{The 3.5--30~keV \nustar\/ light curve in the transitional state during epoch number 2 (top) and in the hard state during epoch number 1 (bottom) using a bin size of 1 s.
\label{fig:lcB}}
\end{center}
\end{figure*}

\begin{figure}[h]
\begin{center}
\includegraphics[angle=-90,width=8.0cm]{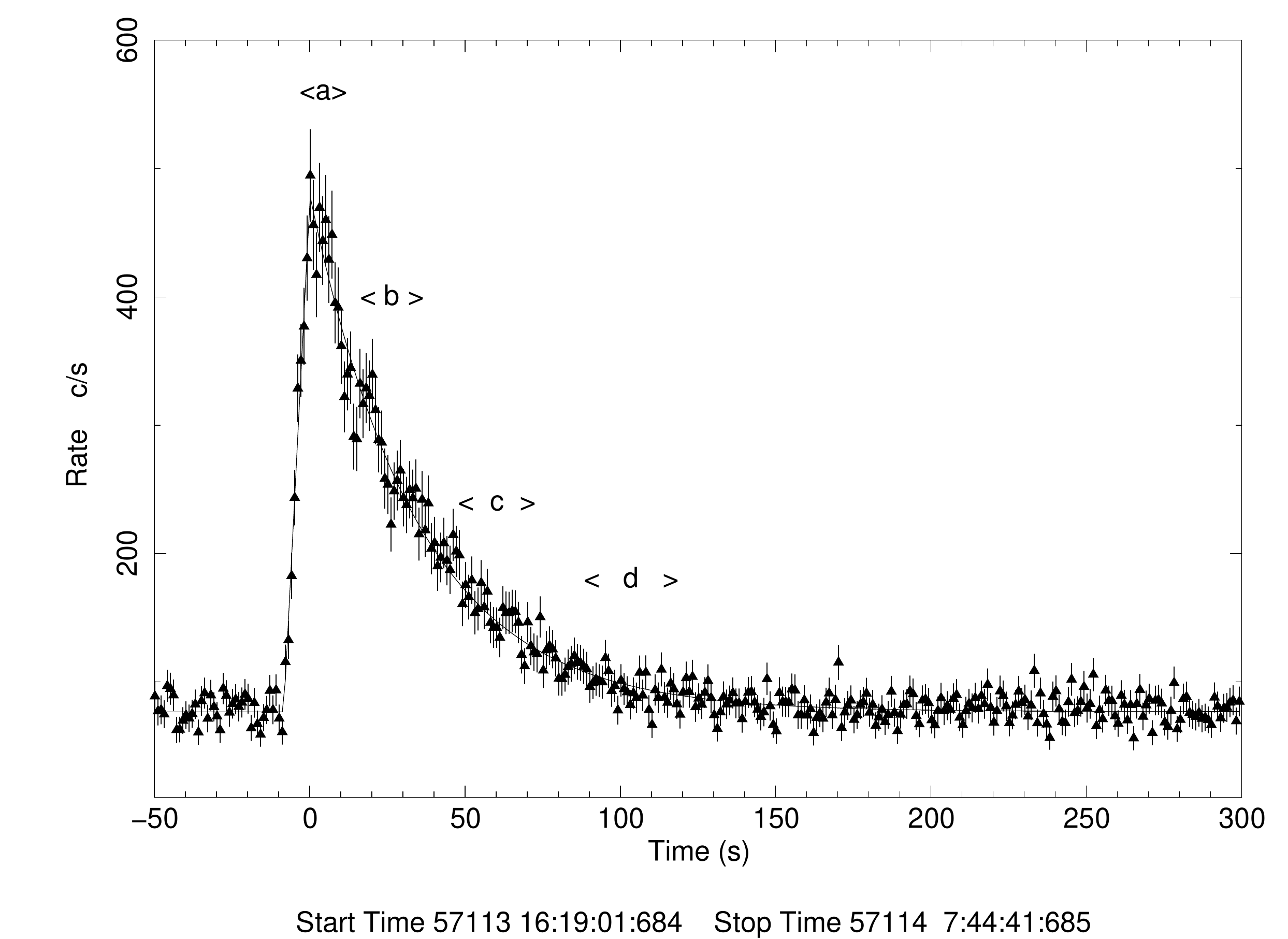}
\includegraphics[angle=-90,width=8.0cm]{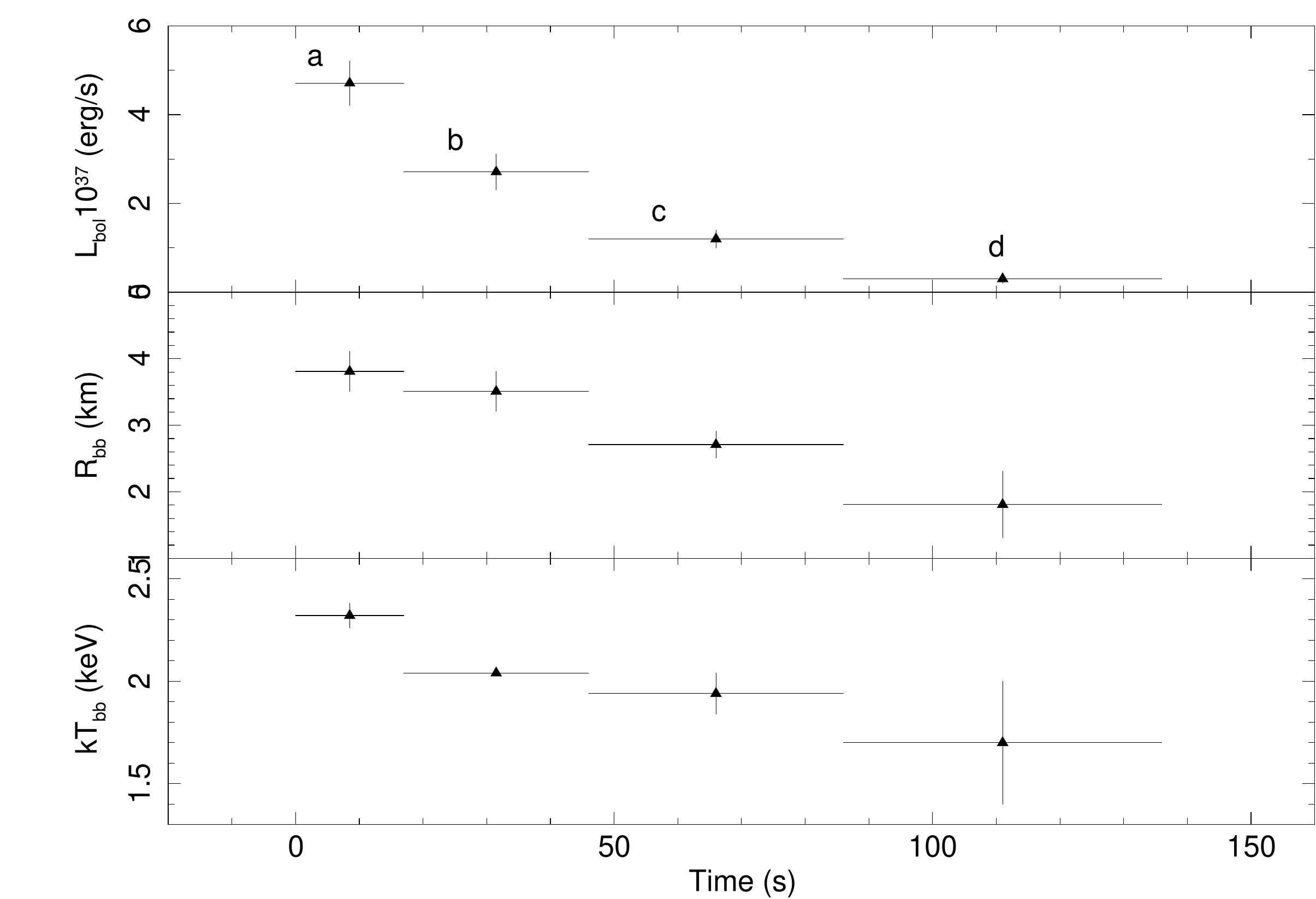}
\caption{Top: profile of the X-ray burst number 5 (zoom of the Fig. \ref{fig:lcB}) in the $3.5-20$~keV energy range with the model consisting of a linear rise followed by an exponential decay overlaid.
Bottom: the change in luminosity, black body radius, and black body temperature (at 5.8 kpc) with time from the time resolved spectral fitting burst number 5. See Table 4 for parameter values.
\label{fig:zoom}}
\end{center}
\end{figure}

We searched for the presence of timing signatures such as Type I X-ray bursts during the observations of \source\/. 
These were detected in a quasi periodic manner during the transitional state, when \nustar\/ observed \source\/ for $\sim\/$ 57 ks. 
The top panel of figure \ref{fig:lcB} shows the light curve with a bin time of 1 s in the 3.5--30~keV energy range. 
 The presence of ten thermonuclear Type-I bursts can be seen clearly within the light curve (see also Wijnands et al.\ 2017). 
The light curve does not show any statistically significant intensity variations outside of the X-ray bursts.    
Unfortunately, there are no data from \integral\//JEM-X or \swift\//XRT during X-ray bursts detected with \nustar\/ instruments. 
To analyze the morphological properties of these bursts, we modeled the burst shape with linear rise ($I=I_{peak}(t-T_{start})/(T_{peak}-T_{start})$, for $T_{start}\leq T \leq T_{peak}$) followed by an exponential decay ($I=I_{peak} e^{-(t-T_{peak})/\tau}$, for $T \geq T_{peak}$), adding a constant factor to take into account the persistent flux.
In Table \ref{tab:burstspec0} we report the start time ($T_{start}$), peak time ($T_{peak}$), exponential decay time ($\tau$), and peak intensity (in unit of c/s) of each X-ray burst. 
All the bursts show the same characteristics: a rise time between $\sim$8 and $\sim$11 s, an exponential decay time between $\sim$33 and $\sim$ 38 s, and a peak intensity between $\sim$ 335 c/s and $\sim$402 c/s. 
As an example we show in  Fig. \ref{fig:zoom} the shape of burst number 5 fit with a linear rise between $T_{start}$ and $T_{peak}$ and an exponential decay after the peak.

To study the spectral characteristics of the Type-I X-ray burst emission, we carried out a time resolved spectral analysis in the 3.5--20~keV energy range. 
We fit each time resolved burst spectrum using a black body model with the pre-burst emission subtracted as background. 
The selected times are marked with symbol a, b, c and d in Fig. \ref{fig:zoom} (left panel). 
Spectral parameters are reported in Table \ref{tab:burstspec} while the time resolved spectroscopic results are shown in  Fig. \ref{fig:zoom} (right panel).

From Figure \ref{fig:lcB} (top panel) it is clear that these X-ray bursts are appearing at regular intervals: for four subsequent events (between 1 to 2,   4 to 5, 7 to 8, 10 to 11 in the top panel of Figure \ref{fig:lcB}) the average waiting time between the bursts is $3990\pm77$ s measured from the burst peak. 
For five bursts (2 to 4, 5 to 7, 8 to 10, 11 to 13 and 13 to 15 of top panel in Figure \ref{fig:lcB})  the average waiting time is $7498\pm130$s, which is about twice the previous waiting time.
Assuming the presence of {\it clocked} X-ray bursts (i.e. source showing repeated X-ray bursts at regular time intervals, see Ubertini et al.\ 1999), we compute the time in which intermediate bursts would have happen (time 3, 6, 9, 12, and 14 of top panel of Figure \ref{fig:lcB}).  We note that, unfortunately, these times occurred during the \nustar\/ orbital data gaps.\\
The $\sim 41$ ks \xmm\/ light curve (Figure 3 of Wijnands et al.\ 2017), starting 4381 s after the \nustar\/ observation (between burst number 1 and number 2 of Figure \ref{fig:lcB}), showed eleven consecutive Type I X-ray bursts every $\sim 3.7 $ks, confirming the {\it clocked} burster behavior highlighted in the \nustar\/ data. 

During the very hard state, \nustar\/ observed \source\/ for $\sim\/$ 80 ks. The bottom panel of Figure \ref{fig:lcB} shows the light curve with a bin time of 1 s in the energy range of 3.5--30~keV.
The light curve indicates the presence of four  thermonuclear Type-I X-ray bursts (number 2, 3, 5 and 8).
We model the light curves of these bursts in the same way as done for the data in the transitional state. 
In Table \ref{tab:burstspec0} we report the start time ($T_{start}$), peak time ($T_{peak}$), exponential decay time ($\tau$), and peak intensity (in unit of c/s) of each X-ray burst during this epoch.  
During this observation the X-ray bursts do not appear {\it clocked} as evident in the transitional state.
 Moreover, the $\sim 57$ ks \xmm\/ light curve (Figure 3 of Wijnands et al.\ 2017), starting about five hours 
after the end of \nustar\/ observation, showed eleven consecutive Type I X-ray bursts every $\sim 8.2$ ks. 
We note that two consecutive Type-I X-ray bursts in the \nustar\/ light curve (number 2 and 3 of the bottom panel of Figure \ref{fig:lcB}) occurred with a waiting time of 7781$\pm$2 s.  
The X-ray bursts number 5 and 8 occurred about twice and three times this value, respectively.
Assuming the presence of {\it clocked} X-ray bursts,  we compute the time  in which bursts would have happen during the first \nustar\/ observation (time 1, 4, 6, 7, 9, and 10 of bottom panel of Figure \ref{fig:lcB}).  
As in the 2$^{nd}$ epoch, these times occurred during the \nustar\/ orbital data gaps. 
The X-ray burst number 10 occurred during a short data gap due to good time interval. In the case of X-ray burst number 7, only a tail of the possible Type-I X-ray burst emission is detected due to an orbital data gap.\\
There were no Type-I X-ray bursts present in the data taken during the soft spectral state.\\

\begin{table*}
\begin{center}
\caption{Results from fitting  \nustar\/ light curve of Type-I X-ray bursts} 
\begin{tabular}{l c c c c }
\hline
Burst number&$T_{start}$    &   $T_{peak}$     &     $\tau$       &    Peak intensity \\
&s    &   s     &     s     &    c/s \\
\hline
&    &   2$^{nd}$ epoch   &          &    \\
1&$3179\pm1  $   &   $  3187\pm1  $  &   $  37.5\pm0.3 $  & $   387\pm12   $   \\
2&$7409\pm1  $   &   $  7418\pm1  $  &   $  35.6\pm1.3 $  & $   391\pm13   $   \\
4&$15307\pm1 $   &   $ 15317\pm1  $  &   $37.8\pm1.4   $  & $ 358\pm12     $   \\
5&$19324\pm1 $   &   $ 19332\pm1  $  &   $ 34.3\pm1.2  $  & $  402\pm14    $   \\
7&$27061\pm1 $   &   $ 27069\pm1  $  &   $ 35.7\pm1.1  $  & $  381\pm13    $   \\
8&$30914\pm1 $   &   $ 30921\pm1  $  &   $ 33.1\pm1.2  $  & $  397\pm13    $   \\
10&$ 38406\pm1$   &   $  38415\pm1 $  &   $  36.0\pm1.4 $  & $  373\pm15    $   \\
11&$ 42264\pm1$   &   $  42272\pm1 $  &   $  34.9\pm1.3 $  & $   396\pm13   $   \\
13&$ 49458\pm1$   &   $  49467\pm1 $  &   $  37.5\pm1.4 $  & $   335\pm13   $   \\
15&$ 56622\pm1$   &   $  56630\pm1 $  &   $  36.7\pm2.7 $  & $  370\pm17    $   \\
\hline
&    &   1$^{st}$ epoch   &          &    \\
2&$14293\pm1$    &$14300\pm1$    &38$\pm2$          &376$\pm15$    \\
3&$22073\pm1$    &22081$\pm1$    &28$\pm2$          &322$\pm15$    \\
5&$38269\pm1$    &38277$\pm1$    &41$\pm2$          &302$\pm13$   \\
8&$62031\pm1$    &62040$\pm1$    &31$\pm2$          &328$\pm18$    \\
\hline
\end{tabular}
\label{tab:burstspec0}
\begin{tablenotes}
\item
 The lightcurves were fit in the 3.5--30~keV energy range using a burst component model (linear rise followed by an exponential decay) for the 2$^{nd}$ and 1$^{st}$ epoch. The burst numbers correspond to those in Figure \ref{fig:lcB}. The constant factor is $C=77.16\pm0.10$ c/s and the $\chi^2_{red}= 1.3$ ($d.o.f=44612$) for the 2$^{nd}$ epoch.
These values are  $C=34\pm3$ c/s and the $\chi^2_{red}=2.3$ ($d.o.f=32106$) for the 1$^{st}$ epoch; this high value of $\chi^2$ is due to the tail of the non-fitted burst number 7.
\end{tablenotes}
\end{center}
\end{table*}

\begin{table*}
\begin{center}
\caption{Results from fitting  \nustar\/ spectral data of X-ray Burst number 5 }.
\begin{threeparttable}
\begin{tabular}{l c c c c c c   }
\hline
Interval Time&Duration&$Constant$  (FPMB)  & $kT_{\mathrm{bb}}$  &$R_{\mathrm{bb}}$  &$L_{\mathrm{2-20keV}}$  &$\chi^2_{\nu}$ (dof)  \\
&s& &keV& km&$10^{37} erg s^{-1}$ &\\
\hline
a \dotfill& 17 &1.0$\pm$0.1               &2.32$\pm$0.06            &3.8$\pm$0.2           & 4.7$\pm$0.5 & 125(137)\\
b \dotfill& 29 &1.0$\pm$0.1               &2.03$\pm$0.09            &3.5$\pm$0.3          & 2.7$\pm$0.4 & 139(145)\\
c \dotfill& 40 &1.0$\pm$0.2               &1.9$\pm$0.1              &2.7$\pm$0.3          & 1.2$\pm$0.2 & 174(174)\\
d \dotfill& 50 &1.0$\pm$0.2               &1.7$\pm$0.3              &1.8$\pm$0.3          & 0.3$\pm$0.1 & 148(148)\\
\hline
\end{tabular}
\label{tab:burstspec}
\begin{tablenotes}
\item
The data were modeled using a simple blackbody model during four time intervals. The shape of the burst profile can be seen in Figure \ref{fig:lcB}. We assumed a distance of 5.8 kpc. 
\end{tablenotes}
\end{threeparttable}
\end{center}
\end{table*}

\section{Discussion}
\subsection{Continuum emission}

The broad band analysis of \source\/ indicates that the source moves from a low/hard state to a transitional state and then to a high/soft states. The X-ray luminosity accordingly increases from $L_{\mathrm{0.8-200~keV}}\simeq1.8\times10^{37}~\mathrm{erg}~\mathrm{s}^{-1}$ 
to  $L_{\mathrm{0.8-200~keV}}\simeq2.6\times10^{37}~\mathrm{erg}~\mathrm{s}^{-1}$, 
assuming an intermediate value of $L_{\mathrm{0.8-200~keV}}\simeq2.2\times10^{37}~\mathrm{erg}~\mathrm{s}^{-1}$ during the transitional state (for a distance of 5.8 kpc).  
When the source was in the very hard state, the {\it INTEGRAL}/IBIS data reveals a very hard emission that cannot be fitted using a single Comptonization model. 
In order to properly describe the data, the spectra required two thermal Comptonization components, each with its own reflection component.
The first thermal component is due to Comptonization of soft seed photons of temperature $\sim1.7$ keV by a plasma with $kT_e\sim10$ keV.
Its relativistic reflection component indicates that the inner disc is viewed at an inclination angle of $31^{\circ}-65^{\circ}$. The Fe abundance is $A_{Fe}\sim 0.5$ solar and the disc appears to be mildly ionized ($log \xi\sim 1.3$). The reflection amplitude is lower than 0.2, indicating a relatively low illumination of the disk by the source of the primary Comptonization continuum.  
The second component is due to Comptonization of soft seed photons of temperature $\leq$ 0.1 keV by a hot plasma with $kT_e\sim35$ keV. The reflection fraction  is very high  $refl\_frac\sim1.7$, indicating a slab-geometry for the corona, where all the thermal radiation must propagate through the corona prior to escaping from the system (Dove et al.\ 1997).  

We estimate the optical depth $\tau$ of the Comptonising region using the relation
$$
\Gamma_{comp} = \left[ \frac{9}{4}+ \frac{1}{\tau (1+\frac{\tau}{3})\left(\frac{kT_e}{m_ec^2}\right)} \right]^{1/2}-\frac{1}{2}
$$ 
(Zdziarski et al.\ 1996), where $m_e$ and $c$ are the  
electron mass and the speed of light, respectively and the
values of $\Gamma_{comp}$ and $kT_e$ are reported in Table \ref{tab:reflspec}. 
Assuming an optically thick corona and spherical emission, we calculate the emission radius of the seed photons $R_0 = 3\times10^4d \;[f_{Compton}/(1+y)]^{1/2} (kT_0)^{-2}$ (in units of km), where $d$ is the source distance
in units of kpc,   $y$ is the Compton parameter  $y=4kT_e max(\tau^2,\tau)/(m_ec^2)$ and
$kT_0$ is the seed photons temperature (in 't Zand et al.\ 1999). 
We obtain an optical depth of $\tau\sim5$ and a radius of the seed photon region of $R_0\sim$ 2 km for the first Comptonization component and  $\tau\sim2$ and  $R_0\sim380$ km  for the second  Comptonization component.
We consider the last value only indicative, being the spherical symmetry disfavoured by the high value of the reflection.
These physical parameters 
indicate the presence of hot seed photons ($kT_{bb}\sim1.7$ keV) close to the neutron star surface ($R_0\sim2$ km) which interact with a hot corona ($kT_e\sim10$ keV, $\tau\sim5$).
During the low/very-hard state, next to these standard components (blue lines in Figure \ref{fig:spe}) there is a second emission component (green lines in Figure \ref{fig:spe}) due to a hot corona ($kT_e\sim35$ keV, $\tau\sim2$) interacting with cold ($kT_{bb}\leq0.1$ keV) seed photons located far from neutron star ($R_{0}\sim380$ km).

During the transitional state the spectral behaviour is very similar to low/hard state, with two Comptonization components. The first one with electron temperature of $kT_e\sim8$ keV, soft seed photons temperature of $\sim1.7$~keV, optical depth of $\tau =6$ and seed radius of $R_0\sim2$ km.  Its relativistic reflection component indicates that the inner disc is viewed at an inclination angle of $25^{\circ}-63^{\circ}$. The Fe abundance is $A_{Fe}\sim 0.5$ solar and the ionization disc is $log \xi\sim 1.9$. The reflection amplitude is $\sim0.3$ indicating a value compatible with a spherical geometry of a compact corona inside an outer accretion disc. The second Comptonization component is due to a  hot corona ($kTe\sim34$ keV, $\tau\sim3$) interacting with cold ($kT_{bb}\leq0.1$ keV) seed photons located farther out from the NS than the first component. 
Assuming a Shakura-Sunyaev disc, we can compute the distance from the central source where the temperature is expected be $\sim 0.1$ keV. Assuming that the observed $kT_{bb}\sim1.7$ keV is the temperature at $R_c\sim10$ km, the relation 
$T(R)=[(3GM_{NS}\dot{M}/8\pi R^3\sigma)\times(1-(R/R_c)^{1/2})]^{1/4}$ gives a distance of $\sim400$~km for a disc location at temperature of 0.1 keV. This value is consistent with the radius of the region of the second seed photon population estimated in the previous paragraph, which could be compatible with the external part of the accretion disc.

When \source\/ is in a high/soft state, the continuum from $0.7-50$~keV is well described by 
a disc black body, a thermal Comptonization component, and its reflection.
We obtain a hydrogen column density of $N_H=(0.3\pm0.1)\times10^{22}\ \mathrm{cm}^{-2}$ 
and an inner disc temperature of $kT_{in}\sim1.2$~keV. 
For the Comptonization component we obtain $\Gamma\sim2.2$ and $kT_{e}\sim2.5$~keV. 
The relativistic reflection component indicates that the inner disc is viewed at an inclination angle of $24^{\circ}-56^{\circ}$. The Fe abundance is consistent with Solar composition ($A_{Fe}\sim 0.8$) and the disc appears to be mildly ionized ($log \xi\sim 1.8$). 
The physical parameters are consistent with values obtained from the \nustar\/ and \swift\/ data reported by Degenaar et al.\ (2016), out of $N_H\sim0.3$, $\Gamma\sim2.2$, and  $kT_{e}\sim2.5$~keV. 
Small differences in column density and $\Gamma$ can be explained by the different soft X-ray observations: the \swift\/ spectrum reported here is an average of the  observations
(number 15 and 16 of Table 1) quasi-simultaneous with \integral\/ data, where the data sample analysed by Degenaar et al.\ (2016) includes observations performed after the \integral\/ ones.   
 Furthermore, the high energy \integral\/ observations provided a better estimate of the electron temperature of the corona and indicate that the observed X-ray emission is dominated by thermal Comptonization with a Maxwellian electron distribution. 
Assuming a distance of 5.8 kpc and a correction factor of 1.7 (Shimura \& Takahara 1995), the inner radius of the accretion disc is $9\pm1$ km. This value is consistent with the inner disc radius calculated from {\scriptsize{RDBLUR}} model $R_{in}=12\pm2$ km and with the expected radius of a neutron star,  suggesting that the accretion disc was truncated near the stellar surface (Done et al.\ 2007). 

Following the same procedure described for the low/hard state, we calculate an optical depth of $\tau\sim10$ and a radius of the seed region $R_0\sim8$ km.
Taking into account this physical parameters we can conclude that during the high/soft state the observed emission  could be due to  the seed photons from an accretion disc and/or neutron star surface and a thick electron corona ($\tau\sim10$) with a temperature of $\sim2.5$ keV.  The position and temperature of the seed photons correspond to the one of the disc photons. 

During the high/soft state, a small fraction of the X-ray photons could be ascribed to jet emission, only detected when the thermal Comptonization is not dominating at high energies. However, the X-ray data presented here do not show any X-ray tail above $\sim50$~keV.

In the low/hard state Baglio et al.\ (2016) reported IR/optical/UV emission from the jet simultaneous to the X-ray data presented here. In spite of this evidence, our data set is well fitted with a thermal Comptonization model  that, in agreement with the energetic considerations of Malzac, Belmont, \& Fabian (2009), favors a thermal model.

To verify whether the observed X-ray emission could be  due to Comptonization of soft seed photons from a non-thermal electron population,
we have fitted the data using a 
hybrid thermal/non-thermal model {\scriptsize{EQPAIR}} in {\scriptsize{XSPEC}} (Coppi 1999).
In this model, the disc/corona system is a spherical hot plasma with the electrons illuminated by soft photons from the accretion disc. At low energies the electron distribution is Maxwellian, while at high energies this distribution is non-thermal. 
A fraction {$l_{nth}/l_{h}$} of the power is ejected in the form of  non-thermal  electrons  rather  than  contributing to  the  thermalized distribution ($l_{th}$). 
This model {\scriptsize{TBABS*(EQPAIR+DISKBB)}} gives a good fit $\chi^2_{\rm red} (d.o.f.)= 1.1 (1594)$ for the soft state, indicating a pure thermal emission $l_{nth}/l_{h}=0$. 
During the transitional and very hard state this model gives a $\chi^2_{\rm red}>1.4$ with significant residuals at high energies. Although 
the radio (Gusinskaia et al.\ 2017) and IR/optical/UV (Baglio et al.\ 2016) emission has been explained by the synchrotron radiation, the X-ray behaviour favours a thermal Comptonization model  to explain the high energy emission. 
At this stage, a small contribution from the non-thermal processes  to the X-ray flux can not be ruled out, but it is not the dominant component in these spectral states.

The system shows two emission components with a different evolution from low/hard to high/soft state: the first one is a standard transition in the framework of the truncated disk model (blue lines in Figure \ref{fig:spe}), while the second one is a new component responsible of the peculiar very-hard emission (green lines in Figure \ref{fig:spe}). 
This peculiarity of the high energy behavior presented here is in agreement with the unusual very hard state observed by Parikh et al.\ (2017) in this source. They suggest that the source was not in a standard low/hard state, rather in a new previously unrecognized state. They compare this spectral behaviour with standard low/hard state for LMXB at the same luminosity and report on the spectral and timing extreme properties: the hardness ratio of the spectra is significantly higher than the standard low/hard state spectra and the noise component in their power density spectra is stronger with very low typical frequencies ($\sim$0.01 Hz).
Our analysis shows that this low/very-hard state is due to two Comptonization components.

Similar behavior was observed by D'A{\'\i} et al.\ (2007) at low accretion rates. They observe that a soft thermal emission from accretion disc and a thermal Comptonization component is unable to fit the broad band spectra of Sco~X-1. Strong residuals in the high energy band required a power law component, which could represent a second thermal Comptonization from a hot plasma or a hybrid thermal/non-thermal Comptonization. The main difference between the case of Sco X-1 and \source\/ is the presence of a high energy curvature ($kT_e\sim 35$ keV) in \source\/ that makes more plausible a thermal emission for this source.

A double Comptonization corona was also observed in 4U~1915$-$05 (but see also Gambino et al., 2019 for a different interpretation) and MAXI~J0556-332 in the soft state (Zhang et al.\ 2014; Sugizaki et al.\ 2013),in GS~1826$-$238 in the low/hard state and in Aql~X-1 in low hard state just before it made a transition into the soft state (Ono et al.\ 2016). In the last two cases, the sources were at the highest luminosity end of the hard state ($L\sim0.1\ L_{\mathrm{Edd}}$), at the same luminosity level as the very hard state of \source\/.  The GS~1826$-$238 spectrum was explained with an emission from a soft standard accretion disc Comptonized by a hot electron cloud and a blackbody emission Comptonized by another hotter electron corona.
Strong similarities are found between those two sources: both are clocked bursters and require a double Comptonization component in their low/hard spectral states.

\subsection{X-ray bursts}
The thermonuclear instabilities on accreting neutron stars give information about the physical and chemical conditions of the binary system.  
In \source\/, ten Type-I X-ray bursts have been detected during the transitional state in the \nustar\/ data, all showing same spectral and timing characteristics. 
The chemical composition of the accreting material will influence the burst properties and can provide a tool to study the chemical composition.
In fact, Type-I X-ray burst theory predicts four different regimes in mass accretion rate ($\dot{M}$) for unstable burning (Fujimoto et al.\ 1981, Fushiki \&\ Lamb 1987; see also Bildsten 1998, 2000, Schatz et al.\ 1999; Peng et al.\ 2007; Cooper \& Narayn 2007). 
During the transitional state, the accretion rate was at $\sim4\times10^{-9}M_{\sun}/yr$, indicating that the X-ray bursts are due to mixed H/He burning triggered by thermally unstable He ignition (4--$11\times 10^{-10}$\,${\rm M}_{\sun}$\,${\rm yr}^{-1} \lesssim \dot{\rm M} \lesssim 2\times10^{-8}$\,${\rm M}_{\sun}$\,${\rm yr}^{-1}$).
The chemical composition can be experimentally confirmed through measurements of the burst duration.
Whenever hydrogen is present in the burning material, the burst duration becomes longer with respect to the pure helium burning ($\tau\geq10$ s). 
This is due to the long series of $\beta$ decays in the r-p process (see e.g.\ Bildsten 1998, 2000).
 During the very hard and transitional states, $\tau$ of the observed X-ray bursts is always longer than $\sim 30$ s, confirming the presence of the H in the burning.

\source\/ exhibits regular $\geq 100$ s long Type-I X-ray bursts recurring at approximately periodic intervals of $\sim$ 3.9 ks and $\sim$ 7.9 ks between two successive X-ray bursts during the 2$^{nd}$ and 1$^{st}$ epochs, respectively (see also Wijnands et al.\ 2017).                                                                     
Quasi periodic bursting has been found in only three other sources: GS~1826-24 (Cocchi et al.\ 2000, Ubertini et al 1999, Cornelisse et al.\ 2003, Galloway et al.\ 2004, Chenevez et al.\ 2016), GS 0836-429 (Aranzana et al.\ 2016) and EXO 1745-248 (Matranga et al.\ 2017). 
Taking into account \nustar\/ and \xmm\/ results (Wijnands et al.\ 2017) for \source\/, the regular bursting regime has been observed during very hard and transitional  states when the persistent count rate changes from $\sim$77 c/s to $\sim$34 c/s. The burst frequency moves from $\sim$ 7.9 ks  to $\sim$ 3.9 ks during these two regimes. 
The persistent luminosity ratio of $\sim2.3$  is roughly similar to the waiting time ratio of $\sim2$.                                                      
The correlation between the waiting time and accretion rate is expected by the standard burst model (see Fujimoto et al.\ 1981, Bildsten 1998), indicating that matter accretes on the neutron star surface in a very stable manner and Type-I X-ray burst occurs when enough material is accumulated to trigger the typical thermonuclear event.\\
No X-ray bursts have been observed in the soft state during the \nustar\/, \swift\/,
and \integral\/ observations. 
This is similar to the behavior observed for GS~1826-24 (Chenevez et al.\ 2015).
The burst properties depend on the spectral states for these sources (Kajava et al.\ 2014, Kuuttila et al.\ 2017).  
Indeed, these authors show that the drivers of the bursting behavior are not only the accretion rate and chemical composition of the accreted material, but also the cooling that is linked to the spectral states.

\section{Conclusions}

We have analyzed the spectral evolution of \source\/ in the $0.8-200$~keV range during its 2015 outburst.
The spectra are well described as the sum of thermal Comptonization and reflection due to illumination of the accretion disc.
During the high/soft state, the blackbody component is due to thermal emission of the accreting disc peaking around 1.2~keV, plus an additional component indicating the Comptonization of disc/NS radiation by a population of electrons with a temperature of $\sim2.5$~keV. 
Accordingly, during the low/very-hard and transitional states, 
the broad band spectra can be explained by two different components:
\begin{itemize}
    \item 
the first thermal Comptonization arises from hot seed photons ($kT_{bb}\sim1.7$ keV) and a cold corona ($kT_e\sim8-10$ keV) close to NS/accretion disc ($R_0\sim2$ km)
\item
the second thermal Comptonization from cold seed photons ($kT_{bb}\leq0.1$ keV) and a hot corona ($kT_e\sim35$ keV) far from the central region ($R_0\sim390$ km). This distance is the expected value for a disc temperature of 0.1 keV, suggesting that the seed photons come from external part of the accretion disc. 
\end{itemize}
The first thermal component indicates a standard transition from a high/soft state to low/hard state in the framework of the truncated disc (Done et al.\ 2007), while the second one explains the peculiar very-hard state.

Finally, \nustar\/ observations show that this source is a new {\it clocked} burster 
with an average waiting time between two successive X-ray bursts of $\sim$ 7.9 ks and $\sim$ 4.0 ks when the persistent luminosity decreases of a factor $\sim2$, moving from the very hard to the transitional  state. The Type-I X-ray burst properties indicate that the thermonuclear emission is due to mixed H/He burning triggered by thermally unstable He ignition.

\acknowledgements
We thank Dr.\ N.\ Degenaar and J.\ van den Eijnden for suggestions and discussion, improving our work.\\
We acknowledge  the ASI financial/programmatic support via contracts ASI-INAF agreement number 2013-025.R1 and ASI-INAF 2017-14-H.0.\\
We acknowledge the use of public data from the \swift\/ and \nustar\/ data archive. \\ 
This research has made use of data provided by 
the High Energy Astrophysics Science Archive Research Center (HEASARC), 
which is a service of the Astrophysics Science Division at NASA/GSFC and the High Energy Astrophysics Division of 
the Smithsonian Astrophysical Observatory.\\
R.L. gratefully acknowledges funding through a NASA Earth and Space Sciences Fellowship and the support of NASA through Hubble Fellowship Program grant HST-HF2-51440.001.\\
F.O. acknowledge the support of the H2020 European HEMERA program, grant agreement No 730970.\\ 


\scriptsize
Aranzana E., Sanchez-Fernandez C. and Kuulkers E., 2016, A\&A, 586,12\\
{Baglio} M.~C., {D'Avanzo} P., {Campana} S., {Goldoni} P., {Masetti} N.,
  {Mu{\~n}oz-Darias} T., {Pati{\~n}o-{\'A}lvarez} V., {Chavushyan} V., 2016,
  \aap, 587, A102\\
Barrow et al., 2005, Space Sci. Rev. 120, 165\\
Bildsten, L. 1998, in NATO Advanced Science Institutes (ASI) Series C, eds. R. Buccheri, J. van Paradijs, and A. Alpar, 515, 419\\
Bildsten L. and Ushomirsky G., 2000, ApJ, 529, 33\\
Boissay et al. 2015,  \atel, 7096, 1B\\
{Chenevez} J. et~al., 2012, \atel, 4050\\
Chenevez J. et al., Astrophys. J.818, 135 (2016)\\
Chiang C., Cackett E. M., Miller J. M., Barret D., Fabian A. C., D'A{\`\i} A., Parker M. L., Bhattacharyya S., Burderi L., Di Salvo T., Egron E., Homan J., Iaria R., Lin D., Miller M. C., 2016, \apj, 821, 105\\
Cocchi M., Bazzano A., Natalucci L., Ubertini P., Heise J., Kuulkers E., in't Zand J. J. M., 2000, AIPC, 510, 203\\
Condon J. J., Cotton W. D., Greisen E. W., Yin Q. F., Perley R. A., Taylor G. B., Broderick J. J., 1998, AJ, 115, 1693\\
Cooper R. L. \& Narayan R.,  2007, ApJ, 661, 468\\
Coppi P. S., 1999, in Poutanen J., Svensson R., eds, ASP Conf. Ser. Vol. 161,
High Energy Processes in Accreting Black Holes. Astron. Soc. Pac., San
Francisco, p. 375\\
Cornelisse et al., 2003, A\&A, 405, 1033\\
Courvoisier T.J.L. et al, 2003, A\&A, 411, 53C\\
{{D'A{\'\i}} A., {{\.Z}ycki} P., {Di Salvo} T., {Iaria} R., 
         {Lavagetto} G. and {Robba}, N.~R.}, ApJ, 667,411\\
{Degenaar} N. et~al., 2015{\natexlab{b}}, \atel, 7352\\
Degenaar N., Altamirano D., Parker M. et al. 2016, MNRAS, 461, 4049\\
Degenaar N., Ballantyne D. R., Belloni T., et al.  2017, High Energy Astrophysical Phenomena, Accepted to Space Science Reviews (arXiv:1711.06272)\\
Di Salvo T., Sanna A., Burderi L., Papitto A., Iaria R., Gambino A. F. and Riggio A., MNRAS, 483, 767\\
Di Salvo T., Iaria R., Matranga M. et~al., 2015, MNRAS, 449, 2794\\
Done Chris, Gierliński Marek and Kubota Aya, 2007, A\&ARv, 15, 1\\
Dove J. B., Wilms J., Maisack M., Begelman M. C., 1997, 487, 759\\
Egron E., Di Salvo T., Motta S. et al., 2013, A\&A, 550, A5 \\
Fabian A. C., Rees, M. J., Stella, L., White N. E., 1989, MNRAS, 238, 729\\
Fabian A.~C., Lohfink A.,  Kara E. et al., 2015, MNRAS, 451, 4375\\
Fiocchi M. et al. 2006, ApJ, 651, 416\\
Fujimoto M. Y., Hanawa T., Miyaji, S., 1981, ApJ, 247, 267\\
Fushiki I. and Lamb D. Q., 1987, ApJ, 323, 55\\
 Galloway D. K. et al, 2004, ApJ, 601, 466\\
 Gambino A. F., Iaria R., Di Salvo T., et al, 2019, A\&A, 625, A92\\
Garcia J. et al. 2014, ApJ, 782, 76\\
{Gehrels} N. et~al., 2004, \apj, 611, 1005\\
Gierlinski  M.,  Zdziarski  A.  A.,  Poutanen  A.  A.,  Coppi  P.,  Ebisawa  K.,
Johnson W. N., 1999, MNRAS, 309, 496 \\
Gusinskaia N. V., et al. 2017, MNRAS in press.\\
{Harrison} F.~A. et~al., 2013, \apj, 770, 103\\
{Iaria} R., Di Salvo T., Del Santo, M et~al., 2016, A\&A, 596, A21\\
 in ’t Zand J.J.M.,  Verbunt F., T.E. Strohmayer, et al., 1999, A\&A, 345, 108 \\
{Kaur} R., {Heinke} C., 2012, \atel, 4085\\
Kajava J.J.E., Nattila J., Latvala O et al. 2014, MNRAS, 445, 4218\\
Kolehmainen M., Done C. \& Diaz Trigo, M. 2011, MNRAS, 416, 311\\
{Krimm} H.~A., {Kennea} J.~A., {Siegel} M.~H., {Sbarufatti} B.,
  2015{\natexlab{a}}, \atel, 7039\\
{Krimm} H.~A. et~al., 2015{\natexlab{b}}, \atel, 6997\\
 Kuuttila J.,  J. J. E. Kajava, J. Nattila, S. E. Motta, C. Sanchez-Fernández, E. Kuulkers, A. Cumming, J. Poutanen, 2017, A\&A in press \\
{Ludlam} R.~M. et~al., 2016, \apj, 824, 37\\
Malzac, Belmont and Fabian, 2009, MNRAS, 400, 1512\\
Matranga M., A. Papitto, T. Di Salvo, E. Bozzo, D. F. Torres,
R. Iaria, L. Burderi, N. Rea, D. de Martino, C, Sanchez-Fernandez,
 A. F. Gambino, C. Ferrigno, and L. Stella, 2017, A\&A, 603, A39\\
Matranga M.,  Di Salvo T.,  Iaria R., et al. 2017, A\&A, 600, A24\\
Markoff, Nowak \& Wilms, 2005, ApJ, 635, 1203\\
Marino A., Del Santo M., Cocchi M., et~al., 2019, MNRAS, in press (arxiv:1909.10359)\\
Mazzola S. M., Iaria R., Di Salvo T., Del Santo M., et~al., 2019, A\&A, 621, A89\\
Migliari S. et al., 2003, MNRAS, 342, 67\\
{Negoro} H. et~al., 2015, \atel, 7008\\
Ono Ko, Sakurai Soki, Zhang Zhongli, Nakazawa Kazuhiro and Makishima Kazuo,
2016, PASJ, 68, 14\\
Parikh A. et~al., 2017a,  MNRAS, 468, 3979\\
Parikh A. et~al., 2017b,  MNRAS, 466, 4074\\
Peng F., Brown E. F., Truran J.W., 2007, ApJ, 654, 1022\\
Pintore F., Di Salvo T., Bozzo E., Sanna A., Burderi L. et~al., 2015, \mnras, 450, 2016\\
Poutaten \& Vurm 2009, ApJ, 690, L97\\
Romano P. et al.,  2006, A\&A, 456, 917\\
Schatz H., Bildsten L., Cumming A., Wiescher M., 1999, ApJ, 524, 1014\\
Shimura, T., \& Takahara, F. 1995, ApJ, 445, 780\\
Sugizaki M. et al. 2013, PAPS, 65, 58\\
 Ubertini et al., 1999, ApJ, 514, 27\\
 Ubertini et al., 2003, A\&A, 411, 131\\
Veledina, Vurm \& Poutanen, 2011, MNRAS, 414, 3330\\
Veledina, Poutanen \& Vurm 2013, MNRAS, 430, 3196\\
Vurm \& Poutanen 2008, IJMPD, 17, 1629 (arXiv:0802.3680)\\
{Voges} W. et~al., 1999, \aap, 349, 389\\
Wijnands et al. 2017, MNRAS, 472, 559\\
Winkler et al. 2003, A\&A, 411, 1\\
Zhang Z., Makishima K., Sakurai S. et, 2014, PAPS, 66,120\\
{Zdziarski} A.~A., {Johnson} W.~N., {Magdziarz} P., 1996, \mnras, 283, 193\\
Zdziarski et al. 2003, MNRAS, 342,355\\
{{\.Z}ycki} P.~T., {Done} C., {Smith} D.~A., 1999, \mnras, 309, 561\\

\end{document}